\documentclass[aps,prd,twocolumn,showpacs,floatfix,eqsecnum]{revtex4} 

\usepackage{graphics}
\usepackage{graphicx}

\usepackage{latexsym}
\usepackage[cp866]{inputenc}

\input epsf

\begin{document}

\title{\boldmath
Mechanisms of the isospin-breaking decay $f_1(1285) \to
f_0(980)\pi^0\to\pi^+\pi^-\pi^0$}
\author{N.~N.~Achasov,$^{1}$\footnote{achasov@math.nsc.ru}
A.~A.~Kozhevnikov,$^{1,2}$\footnote{kozhev@math.nsc.ru} and
G.~N.~Shestakov\,$^{1}$\footnote{shestako@math.nsc.ru}}
\affiliation{$^1$\,Laboratory of  Theoretical Physics,
S.~L.~Sobolev Institute for Mathematics, 630090 Novosibirsk,
Russia,
\\$^2$\,Novosibirsk State University, 630090 Novosibirsk, Russia}


\begin{abstract} Estimated are the contributions of
the following mechanisms responsible for the  decay $f_1(1285)\to
f_0(980)\pi^0\to\pi^+\pi^-\pi^0$: 1) the contribution of the
$a^0_0(980)-f_0(980)$ mixing, $f_1(1285)\to a_0(980)\pi^0\to
(K^+K^-+K^0\bar K^0)\pi^0\to f_0(980)\pi^0\to\pi^+\pi^-\pi^0$, 2)
the contribution of the transition  $f_1(1285)\to(K^+K^-+K^0\bar
K^0)\pi^0\to f_0 (980)\pi^0\to\pi^+\pi^-\pi^0$, arising due to the
pointlike decay $f_1 (1285)\to K\bar K\pi^0$, 3) the contribution of
the transition $f_1(1285)\to(K^*\bar K+\bar K^*K)\to(K^+K^-+K^0\bar
K^0)\pi^0\to f_0(980)\pi^0\to \pi^+\pi^-\pi^0$, where
$K^*=K^*(892)$, and  4) the contribution of the transition
$f_1(1285)\to(K^*_0\bar K+\bar K^*_0K)\to(K^+K^-+K^0\bar
K^0)\pi^0\to f_0(980)\to\pi^+ \pi^-\pi^0$, where $K^*_0=K^*_0(800)$
(or $\kappa$) and $K^*_0(1430)$. These mechanisms break the
conservation of the isospin  due to the nonzero mass difference of
the $K^+$ and $K^0$ mesons. They result in the appearance of the
narrow resonance structure in the $\pi^+\pi^-$ mass spectrum in the
region of the $K\bar K$ thresholds, with the width
$\approx2m_{K^0}-2m_{K^+}\approx8$ MeV. The observation of such a
structure in experiment is the direct indication on the $K\bar K$
loop mechanism of the breaking of the isotopic invariance.  We point
out that existing data should be more precise, and it is difficult
to explain them using the single specific mechanism from those
listed above. Taking the decay $f_1(1285)\to f_0(980)\pi^0
\to\pi^+\pi^-\pi^0$ as the example, we discuss the general approach
to the description of the $K\bar K$ loop mechanism of the breaking
of isotopic invariance.
\end{abstract}


\maketitle

\section{INTRODUCTION}
~

At the end of the 1970s, a threshold phenomenon known as the mixing
of $a^0_0(980)$ and  $f_0(980)$ resonances that breaks the isotopic
invariance was theoretically discovered in Ref. \cite{ADS79}; see
also Ref. \cite{ADS81}. Since that time many new proposals appeared,
concerning both searching it and estimating the effects related with
this phenomenon \cite{Dz,AS97,KerT,CK1,Gr,
AK02,BHS,Ku,Bu1,Ko,Ha,AT,Bu2,WYW,AS04a,AS04b,WZZ07,WZ08,Ro,CK2,
SKum1,SKum2,ADO15,Wa16}. Recently, the results of the first
experiments on its discovery in the reactions   \\[0.4cm]
(a)\ \ \ $\pi^-N\to\pi^-f_1(1285)N\to\pi^-f_0(980)\pi^0N\to$

\vspace*{0.15cm} $\ \ \ \ \to$\,$\pi^-\pi^+\pi^-\pi^0N$\ \ \cite{Do08,Do11}, \\[0.4cm]
(b)\ \ \ $J/\psi\to\phi f_0(980)\to\phi a_0(980)\to\phi\eta\pi^0$\ \ \cite{Ab1}, \\[0.4cm]
(c)\ \ \ $\,\chi_{c1}\to a_0(980)\pi^0\to f_0(980)\pi^0\to\pi^+\pi^-\pi^0$\ \ \cite{Ab1}, \\[0.4cm]
(d)\ \ \ $J/\psi\to\gamma\eta(1405)\to\gamma f_0(980)\pi^0\to\gamma\,3\pi$\ \ \cite{Ab2}, \\[0.4cm]
(e)\ \ \ $J/\psi\to\phi f_0(980)\pi^0\to\phi\,3\pi$\ \ \cite{Ab3}, \\[0.4cm]
(f)\ \ \ $J/\psi\to\phi f_1(1285)\to\phi f_0(980)\pi^0\to\phi\,3\pi$\ \ \cite{Ab3} \\[0.4cm]
have been obtained with the help of detectors VES in Protvino
\cite{Do08,Do11} and BESIII in Beijing \cite{Ab1,Ab2,Ab3}. The
theoretical considerations concerning the BESIII data \cite{Ab2} on
the reaction (d), that is, on the decay $\eta(1405)\to
f_0(980)\pi^0\to3\pi$, were presented in
\cite{WLZZ12,ALQWZ12,WWZZ13,AKS15}.

Interest in the $a^0_0(980)-f_0(980)$ mixing [1--35] is primarily
due to the fact that the amplitude of the isospin breaking
transition $a^0_0(980)$\,$\to $\,$(K^+K^-+K^0\bar K^0)$\,$\to$\,$f_0
(980)$, caused by the mass difference of the $K^+K^-$ and  $K^0\bar
K^0$ intermediate states, in the region between $K\bar K$ thresholds
turns out to be of the order of $\sqrt{(m_{K^0}-m_{K^+})/m_{K^0}}$
[1] [i.e. of the order of the modulus of difference of the phase
space volumes of the $K^+K^-$ and $K^0\bar K^0$ intermediate states:
$|\rho_{K^+K^-}(s)-\rho_{K^0\bar K^0}(s)|$, where $\rho_{K^+K^-}
(s)=\sqrt{1-4m^2_{K^+}/s}$, $\,\rho_{K^0\bar K^0}(s)=\sqrt{1-4
m^2_{K^0}/s}$, $s$ stands for the square the invariant mass of
$K\bar K$ system]), but not $(m_{K^0}-m_{K^+})/m_{ K^0}$, i.e., by
the order of magnitude greater than it could be expected from the
naive considerations. It is natural to expect the relative magnitude
of the isospin violation to be suppressed outside the $K\bar K$
threshold region, i.e., at the level of $(m_{K^0}-m_{K^+})/m_{K^0}$.
To the first approximation, one can neglect these not really
calculable contributions. Thus, in corresponding reactions the
$a^0_0(980)-f_0(980)$ mixing has to manifest itself in the form of
the narrow peaks (with the width of about 10 MeV) in the mass
spectra of the final $\pi^+\pi^-$ or $\eta\pi^0$ mesons.

The narrow resonancelike structure breaking of the isotopic
invariance have been observed in the $\pi^+\pi^-$ and $\eta\pi^0$
mass spectra in all the above reactions (a)--(f). At the same time,
the very large isospin breaking effects discovered in the decays
$f_1(1285)\to f_0(980)\pi^0\to\pi^+\pi^-\pi^0$ and $\eta(1405)\to
f_0(980)\pi^0\to\pi^+\pi^-\pi^0$ in the reactions (a), (d), and (f)
are indicative of the more general $K\bar K$ loop mechanism of the
isospin breaking in these decays. Of course, the data need further
confirmation.

In the present work, we study the mechanisms which could be
responsible for the isospin breaking decay  $f_1(1285)\to
f_0(980)\pi^0\to\pi^+\pi^-\pi^0$. The paper is organized as follows.
The experimental data on this decay are presented in Sec.
\ref{data}. In Sec.~\ref{estimates}, the estimates are given of the
coupling constants squared of the $f_0(980) $ and $a_0(980)$
resonances, $g^2_{f_0\pi^+\pi^-}$, $g^2_{f_0K^+K^-}$,
$g^2_{a_0\eta\pi}$, and  $g^2_{a_0K^+K^-}$, obtained by us using the
data on the intensities of the $a^0_0(980)-f_0(980)$ mixing,
$\xi_{fa}$ and $\xi_{af}$, measured in the reactions (b) and (c),
respectively. These estimates are used in the subsequent analysis.
The contribution of the transition  $f_1(1285)\to a_0(980)\pi^0\to
(K^+K^-+K^0\bar K^0)\pi^0\to f_0(980)\pi^0\to\pi^+\pi^-\pi^0$
arising due to the $a^0_0(980)-f_0(980)$ mixing is discussed in
Sec.~\ref{sec4}. The contributions of the following transitions,
$f_1(1285)\to(K^+K^-+K^0\bar K^0)\pi^0\to f_0(980)\pi^0\to\pi^+
\pi^-\pi^0$ arising due to the pointlike decay $f_1(1285)\to K\bar
K\pi^0$, $f_1(1285)\to(K^*\bar K+\bar K^*K)\to(K^+ K^-+K^0\bar
K^0)\pi^0\to f_0(980)\pi^0\to \pi^+\pi^- \pi^0$, where $K^*=K^*(892
)$, and $f_1(1285)\to(K^*_0\bar K+\bar K^*_0K)\to(K^+K^-+K^0\bar
K^0)\pi^0\to f_0(980)\to\pi^+ \pi^-\pi^0$, where $K^*_0=K^*_0(800)$
(or $\kappa$) and $K^*_0(1430)$, are scrutinized in
Secs.~\ref{sec5}, \ref{sec6}, and \ref{sec7}, respectively. Note
that here we consider the effect of the isospin violation in the
decay $f_1(1285)\to f_0(980)\pi^0\to\pi^+ \pi^-\pi^0$ as being due
solely to the mass difference of the stable charged and neutral $K$
mesons. In Sec.~\ref{sec8}, the general approach to the description
of the $K\bar K$ loop mechanism of the violation of the isotopic
invariance is discussed. Some general comments about our estimates
are given in Sec.~\ref{sec9}. The conclusions concerning the role of
the considered mechanisms of the decay $f_1(1285)\to\pi^+\pi^-\pi^0$
and the discussion of the further studies are presented in
Sec.~\ref{sec10}.


\section{\boldmath THE DATA ON $f_1(1285)\to\pi^+\pi^-\pi^0$}
\label{data} ~

In the VES experiment  \cite{Do11} on reaction (a), the isotopic
symmetry breaking decay $f_1(1285)\to\pi^+\pi^-\pi^0$ was observed,
and the ratio
\begin{eqnarray}\label{EqII-1}
\frac{BR(f_1(1285)\to f_0(980)\pi^0\to\pi^+\pi^-\pi^0)}{
BR(f_1(1285)\to\eta\pi^+\pi^-)}\nonumber
\\ =(0.86\pm0.16\pm0.20)\%.\qquad\quad\
\end{eqnarray}was measured. From this ratio, taking into account the
Particle Data Group (PDG) data \cite{PDG10} on  $BR(f_1
(1285)\to\eta\pi^+\pi^-)$, it was found in Ref.~\cite{Do11} (see
also \cite{PDG14}) that
\begin{eqnarray}\label{EqII-2}
BR(f_1(1285)\to f_0(980)\pi^0\to\pi^+\pi^-\pi^0)\nonumber \\
=(0.30\pm0.09)\%.\qquad\qquad\
\end{eqnarray}
Taking into account the PDG data  \cite{PDG14} on $BR(f_1(1285)\to
a^0_0(980)\pi^0\to\eta\pi^0\pi^0)$, this results in
\begin{eqnarray}\label{EqII-3}
\frac{BR(f_1(1285)\to f_0(980)\pi^0\to\pi^+\pi^-\pi^0)}{
BR(f_1(1285)\to a^0_0(980)\pi^0\to\eta\pi^0\pi^0)}\nonumber
\\ =(2.5\pm0.9)\%.\qquad\quad\qquad
\end{eqnarray}

The decay $f_1(1285)\to\pi^+\pi^-\pi^0$ was also observed in the
BESIII experiment \cite{Ab3} on reaction (f), and the ratio of the
branching fractions
\begin{eqnarray}\label{EqII-4}
\frac{BR(f_1(1285)\to f_0(980)\pi^0\to\pi^+\pi^-\pi^0)}{
BR(f_1(1285)\to a^0_0(980)\pi^0\to\eta\pi^0\pi^0)}\nonumber
\\ =(3.6\pm1.4)\%\qquad\quad\qquad
\end{eqnarray}was obtained.

One more indication on the decay $f_1(1285)/\eta(1295)\to\pi^+\pi^-
\pi^0$ was obtained in the BESIII experiment  \cite{Ab2}, together
with the data on reaction (d). If one attributes it solely to the
$f_1(1285)$ resonance then the following ratio of intensities will
be obtained:
\begin{eqnarray}\label{EqII-5} \frac{BR(f_1(1285)\to
f_0(980)\pi^0\to\pi^+\pi^-\pi^0)}{ BR(f_1(1285)\to
a^0_0(980)\pi^0\to\eta\pi^0\pi^0)}\nonumber
\\ =(1.3\pm0.7)\%.\qquad\quad\qquad
\end{eqnarray}

So, according to the data of the first experiments, the portion of
the isospin-forbidden decay $f_1(1285)\to f_0(980)\pi^0
\to\pi^+\pi^-\pi^0$ relative to the isospin-allowed decay
$f_1(1285)\to a^0_0(980)\pi^0\to\eta\pi^0\pi^0$ could amount to the
quantity from one to four percent. This is large for the quantity
which, at the first sight, could be naturally expected to have the
magnitude  at the level of   $10^{-4}$. The data indicate
undoubtedly on the existence of the mechanisms that enhance the
intensity of the decay $f_1(1285)\to\pi^+\pi^-\pi^0$. Also, the
characteristic feature of this decay is the dominance of the narrow
resonance structure in the $\pi^+\pi^-$ mass spectrum in the
vicinity of the $K\bar K$ thresholds \cite{Do11,Ab3}. Notice that
the enhancement of the decay  $f_1(1285)\to\pi^+\pi^-\pi^0$ and the
narrow structure in the $\pi^+\pi^-$ mass spectrum were expected as
being due to the isospin breaking mechanism of the
$a^0_0(980)-f_0(980)$ mixing \cite{ADS79,ADS81}.

To be specific, when comparing below the theoretical estimates
with the experimental data, we will base our treatment  on the VES
data considering them as average.


\section{\boldmath THE COUPLINGS OF THE $f_0(980)$ AND
$a_0(980)$ FROM THEIR MIXING} \label{estimates} ~

When calculating the $f_1(1285)\to\pi^+\pi^-\pi^0$ decay
probability, we need the values of the coupling constants of the
$f_0(980)$ and $a_0(980)$ resonances with the $\pi\pi$, $K\bar K$,
and $\eta\pi$ channels. Here, we evaluate these coupling constants
using the data on the $a^0_0(980)-f_0(980)$ mixing  \cite{Ab1}. Such
estimation is of interest because earlier it was not realized.

The BESIII collaboration  \cite{Ab1} made the measurements of the
intensity of the $a^0_0(980)-f_0(980)$ mixing in the decays
$J/\psi\to\phi f_0(980)\to\phi a_0(980)\to\phi\eta\pi$ and
$\psi'\to\gamma\chi_{c1}\to\gamma a_0(980)\pi^0\to\gamma
f_0(980)\pi^0\to\gamma\pi^+\pi^-\pi^0$. As a result, the intensities
$\xi_{fa}$ and $\xi_{af}$ of the transitions  $f_0(980)\to a^0_0
(980)$ and $a^0_0(980)\to f_0(980)$, respectively, were obtained:
\begin{eqnarray}\label{EqIII-1}
\xi_{fa}=\frac{BR(J/\psi\to\phi f_0(980)\to\phi a^0_0(980)\to
\phi\eta \pi^0)}{BR(J/\psi\to\phi f_0(980)\to\phi\pi\pi)}\qquad
\nonumber\\ =(0.60\pm0.20(stat.)\pm0.12(sys.)\pm0.26(para.))\%,
\quad \end{eqnarray}
\begin{eqnarray}\label{EqIII-2}
\xi_{af}=\frac{BR(\chi_{c1}\to a^0_0(980)\pi^0\to f_0(980)\pi^0
\to\pi^+\pi^-\pi^0)}{BR(\chi_{c1}\to a^0_0(980)\pi^0\to\eta \pi^0
\pi^0)} \nonumber\\ =(0.31\pm0.16(stat.)\pm0.14(sys.)\pm
0.03(para.))\%. \quad\end{eqnarray} The information concerning the
denominators of Eqs.~(\ref{EqIII-1}) and (\ref{EqIII-2}) was taken
in Ref.~\cite{Ab1} from the works \cite{Ab4} and \cite{PDG10},
respectively. Since the $a^0_0(980)-f_0(980)$ mixing is determined
mainly by the contribution of the $K\bar K$ loops
\cite{ADS79,ADS81,AS04a}, we take in what follows \cite{FN1}
\begin{eqnarray}\label{EqIII-3}
\xi_{fa}=\frac{BR(f_0(980)\to K\bar K\to a^0_0(980)
\to\eta\pi^0)}{BR(f_0(980)\to\pi\pi)},\\
\label{EqIII-3a} \xi_{af}=\frac{BR(a^0_0(980)\to K\bar K\to
f_0(980)\to\pi^+\pi^-)} {BR(a^0_0(980)\to\eta\pi^0)},\end{eqnarray}
where
\begin{eqnarray}\label{EqIII-4}
BR(f_0(980)\to K\bar K\to a^0_0(980)\to\eta\pi^0)=\qquad\ \nonumber\\
\int\left|\frac{\sqrt{s}M_{a^0_0f_0}(s)}{D_{a^0_0}(s)D_{f_0}(s)-s
M^2_{a^0_0f_0}(s)}\right|^2\frac{2s\Gamma_{a^0_0\to\eta\pi^0}(s)}{
\pi}d\sqrt{s},\ \ \end{eqnarray}
\begin{eqnarray}\label{EqIII-5}
BR(a^0_0(980)\to K\bar K\to f_0(980)\to\pi^+\pi^-)=\qquad\ \nonumber\\
\int\left|\frac{\sqrt{s}M_{a^0_0f_0}(s)}{D_{a^0_0}(s)D_{f_0}(s)-s
M^2_{ a^0_0f_0}(s)}\right|^2\frac{2s\Gamma_{f_0\to\pi^+\pi^-}(s)}
{\pi}d\sqrt{s},\ \ \end{eqnarray}
\begin{eqnarray}\label{EqIII-6}
BR(f_0(980)\to\pi\pi)=\int\frac{2s\Gamma_{f_0\to\pi\pi}
(s)}{\pi|D_{f_0}(s)|^2}\,d\sqrt{s},
\end{eqnarray}
\begin{eqnarray}\label{EqIII-7}
BR(a^0_0(980)\to\eta\pi^0)=\int\frac{2s\Gamma_{a^0_0\to\eta\pi^0}
(s)}{\pi|D_{a^0_0}(s)|^2}\,d\sqrt{s}.
\end{eqnarray}In the above expressions,
$D_r(s)$ is the inverse propagator of the unmixed resonance $r$
$[r=a^0_0(980),f_0 (980)]$ with the mass  $m_r$\,,
\begin{eqnarray}\label{EqIII-8}
D_r(s)=m^2_r-s+\sum_{ab}[\mbox{Re}\Pi^{ab}_r(m^2_r)-\Pi^{ab}_r(s)],
\end{eqnarray} $ab=(\eta\pi^0,\,K^+K^-,\,K^0\bar K^0,\,\eta'\pi^0)$ for
$r=a^0_0(980)$ and $ab=(\pi^+\pi^-,\,\pi^0\pi^0,\,K^+K^-,\,K^0\bar
K^0,\,\eta\eta)$ for $r=f_0(980)$; $s$ is the square of the
invariant mass of the system $ab$; $\Pi^{ab}_r(s)$ stands for the
diagonal matrix element of the polarization operator of the
resonance  $r$ corresponding to the contribution of the $ab$
intermediate state \cite{ADS80}; at $s>(m_a+m_b)^2$,
\begin{eqnarray}\label{EqIII-9}
\mbox{Im}\,\Pi^{ab}_r(s)=\sqrt{s}\Gamma_{r\to ab}(s)=\frac{g^2_{r
ab}}{16\pi}\rho_{ab}(s),\end{eqnarray} where $g_{rab}$ is the
coupling constant of $r$ with $ab$,\,
$\rho_{ab}(s)=\sqrt{s-m_{ab}^{(+)\,2}}\,\sqrt{ s-m_{ab}^{(-)
\,2}}\Big/s$ and $m_{ab}^{(\pm)}$\,=\,$m_a\pm m_b$. The expressions
for $\Pi^{ab}_{r}(s)$ in different domains of $s$ are given in
Appendix \ref{appA}. The propagators of the scalar resonances
$1/D_{a^0_0}(s)$ and $1/D_{f_0}(s)$ constructed with taking into
account the finite width corrections [see Eqs.~(\ref{EqIII-8}),
(\ref{EqIII-9}), (\ref{Eq1A})--(\ref{Eq4A})] satisfy the
K\"{a}ll\'{e}n-Lehman representation and, due to this fact, provide
the normalization of the total decay probability to unity:
$\sum_{ab}BR(r\to ab)=1$ \cite{AKi04}.

The amplitude of $a^0_0(980)-f_0(980)$ mixing,
$\sqrt{s}M_{a^0_0f_0}(s)$, in Eqs.~(\ref{EqIII-4}) and
(\ref{EqIII-5}) is determined by the sum of the one-loop diagrams
$a^0_0(980)\to K^+K^-\to f_0(980)$ and $a^0_0(98 0)\to K^0\bar
K^0\to f_0(980)$ and, by taking into account the isotopic symmetry
for the coupling constants, can be written in the form
\cite{ADS79,ADS81,AS04a}
\begin{widetext}\begin{eqnarray}\label{EqIII-11}
\sqrt{s}M_{a^0_0f_0}(s)=\frac{g_{a^0_0K^+K^-}g_{f_0K^+K^-}}{16\pi}
\left[i[\rho_{K^+K^-}(s)-\rho_{K^0\bar K^0}(s)]-\frac{\rho_{K^+K^-}
(s)}{\pi}\,\ln\frac{1+\rho_{K^+K^-}(s)}{1-\rho_{K^+K^-}(s)}\right.
\nonumber \\ \left.+ \frac{\rho_{K^0\bar K^0}(s)}{\pi}\,\ln\frac{1+
\rho_{K^0\bar K^0}(s)} {1-\rho_{K^0\bar K^0}(s)}\right], &&
\end{eqnarray}\end{widetext} where  $\rho_{K\bar K}(s)=\sqrt{1-4m^2_K
/s}$ at  $\sqrt{s}\geq2m_K$; if  $\sqrt{s}\leq2m_K$ then
$\rho_{K\bar K}(s)$ should be replaced by  $i|\rho_{K\bar K}(s)|$.
In the energy domain with the width of about 8 MeV between $K^+K^-$
and $K^0\bar K^0$ thresholds one has
\begin{eqnarray}\label{EqIII-12}
\left|\sqrt{s}M_{a^0_0f_0}(s)\right|\approx\left|\frac{g_{a^0_0K^+
K^-}g_{f_0K^+K^-}}{16\pi}\right|\sqrt{\frac{m^2_{K^0}-m^2_{K^+}}{m^2_{
K^0}}}\ \nonumber\\ \approx0.127\left|\frac{g_{a^0_0K^+K^-
}g_{f_0K^+K^- }}{16\pi}\right|. \qquad\qquad\end{eqnarray} When
$\sqrt{s}>2m_{K^0}$ and when  $\sqrt{s}<2m_{K^+}$ the quantity
$|\sqrt{s}M_{a^0_0f_0}(s)|$ drops sharply, so that the integrands in
Eqs.~(\ref{EqIII-4}) and ( \ref{EqIII-5}) become the narrow
resonance peaks located near the $K\bar K$ thresholds.

Upon inserting  the central values of $\xi_{fa}$ and $\xi_{af}$ from
Eqs.~(\ref{EqIII-1}) and (\ref{EqIII-2}), respectively, to the
left-hand side of the expressions (\ref{EqIII-3}) and
(\ref{EqIII-3a}) one obtains some equations for the coupling
constants of the $a^0_0(980)$ and $f_0(980)$ resonances which can be
solved numerically. When doing this in such a way, we have obtained
the following estimates:
\begin{eqnarray}\label{EqIII-13a}
\frac{g^2_{f_0\pi\pi}}{16\pi}\equiv\frac{3}{2}\frac{g^2_{f_0\pi^+\pi^-}}
{16\pi}=0.098\mbox{\ GeV}^2,\\ \label{EqIII-13b} \frac{g^2_{f_0
K\bar K}}{16\pi}\equiv2\frac{g^2_{f_0 K^+K^-}}{16\pi}=0.4\mbox{\
GeV}^2, \\ \label{EqIII-13c}
\frac{g^2_{a^0_0\eta\pi^0}}{16\pi}=0.2\mbox{\ GeV}^2, \quad\quad\  \\
\label{EqIII-13d} \frac{g^2_{a^0_0 K\bar
K}}{16\pi}\equiv2\frac{g^2_{a^0_0 K^+K^-}}{16\pi}=0.5\mbox{\ GeV}^2.
\end{eqnarray}
When so doing, we fix the masses of the $a^0_0$ and  $f_0$
resonances to be $m_{a^0_0}=0.985$ GeV and $m_{f_0}=0.985$ GeV,
while the relations of the $q^2\bar q^2$ model,
$g^2_{a^0_0\eta'\pi^0}=g^2_{a^0_0\eta\pi^0}$ and
$g^2_{f_0\eta\eta}=g^2_{f_0 K^+K^-}$, see, e.g., Refs.~\cite{ADS81,
AI89}, are invoked for the estimates of their couplings with the
$\eta'\pi^0$ and $\eta\eta$ channels. The integration  in
Eqs.~(\ref{EqIII-4}) and (\ref{EqIII-5}) is made over the region
from 0.9 to 1.05 GeV, while the corresponding integration interval
in Eqs.~(\ref{EqIII-6}) and (\ref{EqIII-7}) is from the $\pi\pi$ and
$\eta\pi^0$ thresholds, respectively, to 1.3 GeV.

\begin{figure} 
\includegraphics[width=6.5cm]{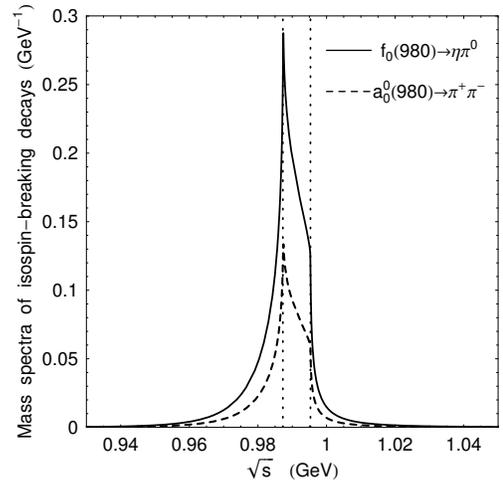}
\caption{\label{Fig1} Mass spectra in the isospin-violating decays
$f_0(980)\to\eta\pi^0$ and $a_0(980)\to\pi^+\pi^-$, caused by the
$a^0_0(980)-f_0(980)$ mixing. The solid and dashed lines are
generally similar each other. The dotted vertical lines show the
locations of the $K^+K^-$ and  $K^0\bar K^0$ thresholds.}
\end{figure}

\begin{figure} 
\includegraphics[width=6.5cm]{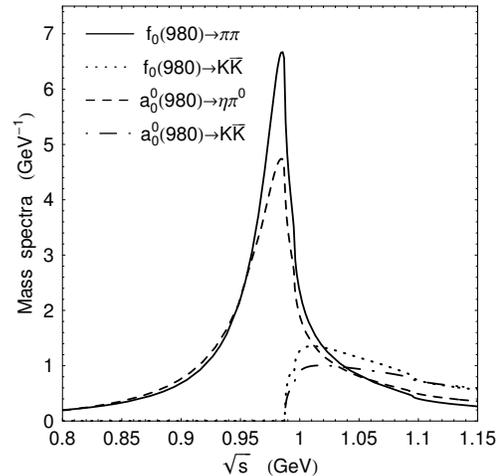}
\caption{\label{Fig2} Mass spectra in the isospin conserving decays
of the $f_0(980)$ and $a^0_0(980)$resonances.}
\end{figure}

Using Eqs.~(\ref{EqIII-12}), (\ref{EqIII-13b}), and
(\ref{EqIII-13d}), one obtains that the ``mass'' of the
$a_0(980)-f_0(980)$ transition $|M_{a^0_0f_0} (4 m^2_K)|\approx28$
MeV. The mass spectra for the isospin-violating and
isospin-conserving decays of the   $f_0(980) $ and $a^0_0(980)$
resonances evaluated as the functions of $\sqrt{s}$ at the earlier
found magnitudes of coupling constants are shown in Figs.~\ref{Fig1}
and \ref{Fig2}. The curves in Fig.~\ref{Fig1} correspond to the
integrands in Eqs.~(\ref{EqIII-4}) and (\ref{EqIII-5}). The curves
in Fig.~ \ref{Fig2} correspond to the integrands in
Eqs.~(\ref{EqIII-6}) and (\ref{EqIII-7}), and to the analogous
expressions for the mass spectra of the decays into $K\bar K$. The
shown spectra look rather usual.

There are a sizable number of works devoted to the evaluation,
estimation, and determination from the fits of the square of the
coupling constants of the $f_0(980)$ and  $a_0(980)$ resonances with
the $\pi\pi$, $K\bar K$ and $\eta\pi$ channels, see, e.g.,
Refs.~\cite{ADS79,ADS81,WZZ07,WZ08,ADS80,AKi04,Fla76,Jaf77,MOS77,
AI89,AG97,AG01,AKi12} (this list does not pretend on completeness).
The spectrum of their possible values is rather wide, so that the
magnitudes of couplings determined from different reactions by
different methods agree within the factor of 2 or greater.  The
values (\ref{EqIII-13a})--(\ref{EqIII-13d}) occupy some average
position among those cited in the literature, so it seems to us to
be natural to use them as the guide in the subsequent analysis.

\begin{figure} 
\includegraphics[width=5.5cm]{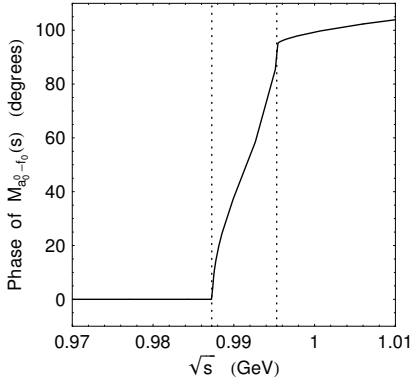}
\caption{\label{Fig1a} The phase of the $a^0_0(980)-f_0(980)
$ mixing amplitude  $M_{a^0_0f_0}(s)$ [See Eq.~(\ref{EqIII-11})].}
\end{figure}

Notice that the phase of the amplitude of the  $a^0_0(980)-f_0(980)$
mixing, $M_{a^0_0f_0}(s)$, in the region between  $K^+K^-$ and
$K^0\bar K^0$ thresholds changes by about $90^\circ$ [see
Eq.~(\ref{EqIII-11}) and Fig.~\ref{Fig1a})].
This fact is crucial for the observation of the
$a^0_0(980)-f_0(980)$ mixing effect in polarization experiments
\cite{AS04a,AS04b}. A similar sharp and large variation of the phase
of the amplitude $f_1(1285)\to(K^+K^-+K^0\bar K^0)\pi^0\to f_0(980)
\pi^0$ in the $f_0(980)$ channel takes place for all mechanisms of
the decay $f_1(1285)\to f_0(980)\pi^0\to\pi^+\pi^-\pi^0$ considered
below. This fact should be also taken into account in suitable
polarization and interference experiments.


\section{\boldmath $a^0_0(980)-
f_0(980)$ MIXING IN THE $f_1(1285)\to\pi^+\pi^-\pi^0$ DECAY}
\label{sec4} ~

Let us calculate the widths of the decays  $f_1(1285)\to
a^0_0(980)\pi^0\to\eta\pi^0\pi^0$, $f_1(1285)\to a_0(980)\pi\to
K\bar K\pi$ and the width of the decay $f_1(1285)\to a^0_0(980)
\pi^0\to f_0 (980)\pi^0\to\pi^+\pi^-\pi^0$ caused by the
$a^0_0(980)-f_0(980)$ mixing \cite{FN2}.

Let us write the  $f_1(1285)\to a^0_0(980)\pi^0\to\eta\pi^0
\pi^0$ decay width in the form
\begin{eqnarray}\label{EqIV-1}
\Gamma_{f_1\to a^0_0\pi^0\to\eta\pi^0\pi^0}=\frac{1}{3}\,\Gamma_{
f_1\to a_0\pi\to\eta\pi\pi}\qquad\nonumber \\ =\frac{g^2_{f_1
a^0_0\pi^0}g^2_{ a^0_0\eta \pi^0}}{192\,\pi^3\,m^3_{f_1}}
\int\limits_{(m_\eta+m_\pi)^2}^{(m_{f_1}-m_\pi)^2}ds
\int\limits_{a_-(s)}^{a_+(s)}dt\ \mathcal{T}(s,t)\,,
\end{eqnarray} where
\begin{eqnarray}\label{EqIV-2}
\mathcal{T}(s,t)=\frac{p^2(s)}{|D_{a^0_0}(s)|^2}+
\mbox{Re}\,\frac{p(s)\,p(t)\,\cos\theta}{
D_{a^0_0}(s)D^*_{a^0_0}(t)}\,,\\
\label{EqIV-3}
a_\pm(s)=\frac{1}{2}(m^2_{f_1}+m^2_\eta+2m^2_\pi-s)\qquad\ \nonumber\\
+\frac{(m^2_{f_1}-m^2_\pi)(m^2_\eta-m^2_\pi)}{2s}
\pm\frac{2m_{f_1}}{\sqrt{s}}\,p(s)q(s)\,,
\end{eqnarray}
\begin{eqnarray}
\label{EqIV-4}
p(s)=\sqrt{m^4_{f_1}-2m^2_{f_1}(s+m^2_\pi)+(s-m^2_\pi)^2}\Bigl/(2m_{f_1}),\ \ \\
\label{EqIV-5}
p(t)=\sqrt{m^4_{f_1}-2m^2_{f_1}(t+m^2_\pi)+(t-m^2_\pi)^2}\Bigl/(2m_{f_1}),\ \ \\
\label{EqIV-6}
q(s)=\sqrt{s^2-2s(m^2_\eta+m^2_\pi)+(m^2_\eta-m^2_\pi)^2}\Bigl/(2\sqrt{s}),\
\ \\ \label{EqIV-7}
p(s)\,p(t)\,\cos\theta=\frac{1}{2}(s+t-m^2_{f_1}-m^2_\eta)\qquad\qquad\nonumber \\
+\frac{(m^2_{f_1}+m^2_\pi-s) (m^2_{f_1}+m^2_\pi-t)}{4m^2_{f_1}}
\,,\qquad\qquad
\end{eqnarray}
$s$ is the square of the invariant mass of the state $\eta\pi^0_1$
and $t$ stands for the square of the invariant mass of the state
$\eta\pi^0_2$ in the decay $f_1(1285)\to\eta\pi^0_1 \pi^0_2$. The
expression $V_{f_1a^0_0\pi^0 }=g_{ f_1
a^0_0\pi^0}(\epsilon_{f_1},p_{\pi^0}-p_{a^0_0})$ is used for the
effective vertex of the  $f_1(128 5)a^0_0(980)\pi^0$ interaction,
where $\epsilon_{f_1}$ is the four-vector of the  $f_1(1285)$
polarization while $p_{\pi^0}$ and  $p_{a^0_0}$ are the four-momenta
of $\pi^0$ and  $a^0_0(980)$, respectively.

The width of the  $f_1(1285)\to a_0(980)\pi^0\to K\bar K\pi$ decay
in the approximation of isotopic symmetry is written in the
following form:
\begin{eqnarray}\label{EqIV-8}\Gamma_{f_1\to a_0\pi\to K\bar K\pi}
=6\,\Gamma_{f_1\to a^0_0\pi^0\to K^+K^-\pi^0}\ \ \  \nonumber \\
=\frac{g^2_{f_1 a^0_0\pi^0}}{\pi\,m^2_{f_1}}
\int\limits_{2m_{K^+}}^{
m_{f_1}-m_{\pi^0}}p^3(s)\,\frac{2s\Gamma_{a^0_0\to K^+K^-}(s)}{
\pi|D_{a^0_0}(s)|^2}\,d\sqrt{s}\,.
\end{eqnarray}
The width of the decay $f_1(1285)\to a^0_0(980)\pi^0\to f_0
(980)\pi^0\to\pi^+\pi^-\pi^0$ caused by the $a^0_0(980)-f_0(980)$
mixing is represented by the expression
\begin{eqnarray}\label{EqIV-9}\Gamma_{f_1\to a^0_0\pi^0\to f_0
\pi^0\to\pi^+\pi^-\pi^0}\qquad\qquad\  \nonumber \\
=\frac{g^2_{f_1a^0_0\pi^0}}{6\pi\,m^2_{f_1}}\int\limits_{0.9
\mbox{\scriptsize{ GeV}}}^{1.05\mbox{\scriptsize{ GeV}}}
\left|\frac{\sqrt{s}M_{a^0_0f_0}(s)}{D_{a^0_0}(s)D_{f_0}(s)-s M^2_{
a^0_0f_0}(s)}\right|^2\nonumber \\
\times\,p^3(s)\,\frac{2s\Gamma_{f_0\to\pi^+
\pi^-}(s)}{\pi}\,d\sqrt{s}\,.\qquad
\end{eqnarray}
The form of $\pi^+\pi^-$ mass spectrum given by the integrand in
Eq.~(\ref{EqIV-9}) is practically indistinguishable from the curves
shown in Fig.~\ref{Fig1}.

As a result of numerical integration Eqs.~(\ref{EqIV-1}),
(\ref{EqIV-8}), and (\ref{EqIV-9}) we obtain
\begin{eqnarray}\label{EqIV-10}
\frac{\Gamma_{f_1\to a^0_0\pi^0\to f_0 \pi^0\to\pi^+\pi^-
\pi^0}}{\Gamma_{f_1\to a^0_0\pi^0\to\eta\ \pi^0\pi^0}}
\qquad\qquad\quad  \nonumber\\ =\frac{BR(f_1\to a^0_0(980)\pi^0\to
f_0(980)\pi^0\to\pi^+\pi^-\pi^0)}{BR(f_1\to a^0_0(980)\pi^0\to
\eta\pi^0\pi^0)}\nonumber\\ \approx0.29\%\,, \qquad\qquad
\qquad\qquad \end{eqnarray}
\begin{eqnarray}\label{EqIV-11}\frac{\Gamma_{f_1\to a_0\pi\to K\bar
K\pi}}{\Gamma_{f_1\to a_0\pi\to\eta\pi\pi}}\qquad\qquad\quad\quad
\nonumber\\ =\frac{BR(f_1\to a_0(980)\pi\to K\bar K\pi)}{BR(f_1\to
a_0(980)\pi \to\eta\pi\pi)}\approx0.11\,.
\end{eqnarray}
The magnitude of the ratio (\ref{EqIV-10}) is close to the central
value of $\xi_{af}$ from Eq.~(\ref{EqIII-2}) but approximately by an
order of magnitude lower than that resulting from the VES data, see
Eq.~(\ref{EqII-3}); see also Eqs.~(\ref{EqII-4}) and (\ref{EqII-5}).
Hence, it is difficult to explain the  VES data by the
$a_0^0(980)-f_0(980)$ mixing mechanism only. In due turn, the PDG
data \cite{PDG14} for the ratio
\begin{eqnarray}\label{EqIV-12}\frac{BR(f_1\to K\bar K\pi)}{BR(f_1
\to a_0(980)\pi \to\eta\pi\pi)}\approx0.25\pm0.05 \end{eqnarray} do
not contradict Eq.~(\ref{EqIV-11}) but indicate that the mechanism
$f_1(1285)\to a_0(980)\pi\to K\bar K\pi$ could be a nonunique source
of the decay $f_1(1285)\to K\bar K\pi$.

It is interesting to reveal at what coupling constants of the
$a^0_0(980)$ and $f_0(980)$ resonances the ratio
\begin{eqnarray}\label{EqIV-10a}
\frac{BR(f_1\to a^0_0(980)\pi^0\to f_0(980)\pi^0\to\pi^+\pi^-\pi^0)
}{BR(f_1\to a^0_0(980)\pi^0\to \eta\pi^0\pi^0)},\nonumber
\end{eqnarray}
calculated for the  $a^0_0(980)-f_0(980)$ mixing mechanism, can be
compatible with the VES data shown in Eq.~(\ref{EqII-3}), i.e.,
$\approx0.025$. Using Eqs.~(\ref{EqIII-8}), (\ref{EqIII-9}),
(\ref{EqIV-1})--(\ref{EqIV-7}), and (\ref{EqIV-9}) we find that the
relation
\begin{equation}\label{EqIV-10b} \frac{BR(f_1\to a^0_0(980)\pi^0\to
f_0(980)\pi^0\to\pi^+\pi^-\pi^0) }{BR(f_1\to a^0_0(980)\pi^0\to
\eta\pi^0\pi^0)}=0.025
\end{equation} is fulfilled if
\begin{eqnarray}\label{EqIV-13a}
\frac{g^2_{f_0\pi\pi}}{16\pi}\equiv\frac{3}{2}\frac{g^2_{f_0\pi^+\pi^-}}
{16\pi}=0.46\mbox{\ GeV}^2,\\ \label{EqIV-13b} \frac{g^2_{f_0 K\bar
K}}{16\pi}\equiv2\frac{g^2_{f_0 K^+K^-}}{16\pi}=2.87\mbox{\ GeV}^2,
\\ \label{EqIV-13c}
\frac{g^2_{a^0_0\eta\pi^0}}{16\pi}=0.48\mbox{\ GeV}^2, \quad\quad\  \\
\label{EqIV-13d} \frac{g^2_{a^0_0 K\bar
K}}{16\pi}\equiv2\frac{g^2_{a^0_0 K^+K^-}}{16\pi}=4.97\mbox{\
GeV}^2,\end{eqnarray}i.e., at rather exotic (large) values of the
coupling constants. The values  of $g^2_{f_0 K\bar K}/(16\pi)$ and
$g^2_{a^0_0 K\bar K}/(16\pi)$ in Eqs.~(\ref{EqIV-13b}) and
(\ref{EqIV-13d}) are by the factors of 7 and 10 greater than in
Eqs.~ (\ref{EqIII-13b}) and (\ref{EqIII-13d}), respectively.  In
this connection, the parameter $\xi_{af}$, evaluated according to
Eq.~(\ref{EqIII-3a}), turns out to be  9 times larger than its
central experimental value in Eq.~(\ref{EqIII-2}). Due to the very
strong coupling of $a^0_0(980)$ with the $K\bar K$ channel, the
width of the $a^0_0(980)$ peak in the $\eta\pi^0$ mass spectrum
turns out to be near 15 MeV in all, and $BR(a^0_0(980)\to\eta\pi^0)$
evaluated over the interval from the $\eta\pi^0 $ threshold to 1.3
GeV, reduces to the magnitude of 6.5\% only. The magnitude of
$BR(f_0(980)\to\pi\pi)$ evaluated over the region from the  $\pi\pi
$ threshold to  1.3 GeV reduces to approximately 12\%.

Since the experimental situation is far from being clear, these
estimates, despite the obtained not-too-satisfactory resonance
characteristics, allow one  to guess the possible role of the
$a^0_0(980)-f_0(980)$ mixing mechanism in the decay $f_1(1285)\to
f_0(980)\pi^0\to\pi^+\pi^-\pi^0$. In what follows, we will not base
our considerations on the values (\ref{EqIV-13a})--(\ref{EqIV-13d}).


\section{\boldmath POINTLIKE DECAY $f_1(1285)\to K\bar K\pi$}
\label{sec5} ~

Let us consider the pointlike mechanism of the  $f_1(1285)$ decay
into the $\pi$ meson and the $S$ wave  $K\bar K$ system. Let us
write the corresponding effective vertex of the
$f_1(1285)K^+K^-\pi^0$ interaction in the form
$V_{f_1K^+K^-\pi^0}=g_{f_1K^+ K^-\pi^0}(\epsilon_{f_1},p_{\pi^0})$.
One has, assuming the isotopic symmetry
\begin{eqnarray}\label{EqV-1}\Gamma_{f_1\to K\bar
K\pi}=6\,\Gamma_{f_1\to K^+K^-\pi^0}\qquad\qquad \nonumber \\
=\frac{g^2_{f_1K^+ K^-\pi^0}}{4\pi}\int\limits_{2m_{K^+}}^{
m_{f_1}-m_{\pi^0}}\frac{p^3(s)\,\rho_{K^+K^-}(s)}{
16\pi m^2_{f_1}}\frac{2\sqrt{s}}{\pi}\,d\sqrt{s}\nonumber \\
=\frac{g^2_{f_1K^+ K^-\pi^0}}{4\pi}\,\times1.46\times10^{-6}\mbox{\
GeV}^3\,.\qquad \end{eqnarray} For the width of the isospin-breaking
transition  $f_1(1285)\to(K^+K^-+K^0\bar K^0)\pi^0\to
f_0(980)\pi^0\to\pi^+\pi^-\pi^0$ caused by the pointlike decay
$f_1(1285)\to K\bar K\pi^0 $ one gets [see Eq.~(\ref{EqIII-11})] the
expression
\begin{eqnarray}\label{EqV-2}\Gamma_{f_1\to(K^+K^-+K^0\bar K^0)
\pi^0\to f_0\pi^0\to\pi^+\pi^-\pi^0}\ \ \ \nonumber \\
=\frac{g^2_{f_1K^+ K^-\pi^0}}{4\pi}\,\frac{1}{6}\,\int\limits_{0.9
\mbox{\scriptsize{ GeV}}}^{1.05\mbox{\scriptsize{ GeV}}}
\left|\frac{\sqrt{s}M_{a^0_0f_0}(s)}{g_{a^0_0K^+K^-}}\right|^2\nonumber \\
\times\,\frac{p^3(s)}{m^2_{f_1}}\,\frac{2s\Gamma_{f_0\to\pi^+
\pi^-}(s)}{\pi|D_{f_0}(s)|^2}\,d\sqrt{s}\qquad\nonumber \\
=\frac{g^2_{f_1K^+ K^-\pi^0}}{4\pi}\,\times3.28\times10^{-9}\mbox{\
GeV}^3\,.\ \ \end{eqnarray} The comparison of Eq.~(\ref{EqV-2}) with
(\ref{EqV-1}) gives
\begin{eqnarray}\label{EqV-3}\frac{\Gamma_{f_1\to(K^+K^-+K^0\bar K^0)
\pi^0\to f_0\pi^0\to\pi^+\pi^-\pi^0}}{\Gamma_{f_1\to K\bar K \pi}}
\nonumber \\ =0.224\times10^{-2}\,.\qquad\qquad\end{eqnarray} This
value is by approximately 15 times lower than the corresponding
central experimental value
\begin{eqnarray}\label{EqV-4}\frac{BR(f_1(1285)\to f_0(980)\pi^0
\to\pi^+\pi^-\pi^0)}{BR(f_1(1285)\to K\bar K \pi)}\nonumber
\\ =0.033\pm0.010\ \qquad\qquad\ \ \end{eqnarray} resulted from
the VES \cite{Do11}, see Eq.~(\ref{EqII-2}), and PDG \cite{PDG14}
data.

The $\pi^+\pi^-$ mass spectrum in the decay $f_1(1285)\to(K^+
K^-+K^0\bar K^0)\pi^0\to f_0(980)\pi^0\to\pi^+\pi^-\pi^0$ whose
expression is given by the integrand in Eq.~(\ref{EqV-2}), looks
similar to the curves in Fig.~\ref{Fig1}. However, it is clear that
the pointlike mechanism of the decay $f_1(1285)\to K\bar K\pi$
cannot by itself provide the considerable probability of the
$f_1(1285)\to f_0(980)\pi^0\to\pi^+\pi^-\pi^0$ transition.


\section{\boldmath DECAY $f_1(1285)\to(K^*\bar K+\bar K^*K)\to
(K^+K^-+K^0\bar K^0)\pi^0\to f_0(980)\pi^0\to\pi^+\pi^-\pi^0$}
\label{sec6}

\begin{figure} [!ht] 
\includegraphics[width=7.6cm]{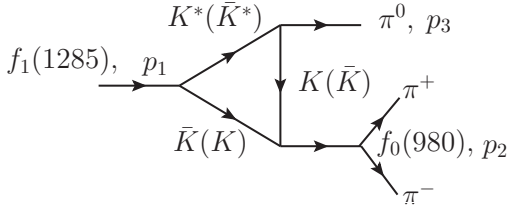}
\caption{\label{Fig3} The diagram of the decay $f_1(1285)\to
f_0(980)\pi^0\to\pi^+\pi^-\pi^0$ via the $K^*\bar K+\bar K^*K$
intermediate states; $p_1$, $p_2$, $p_3$ stand for the four-momenta
of particles participating in the reaction, $p_1^2=m^2_{f_1}$,
$p_2^2=s=m^2_{\pi^+\pi^-}$ is the invariant mass squared of the
$f_0(980)$ or of the final $\pi^+\pi^-$ system, $p_3^2=m^2_{\pi^0}
$.}\end{figure}

\begin{figure} [!ht] 
\includegraphics[width=6.4cm]{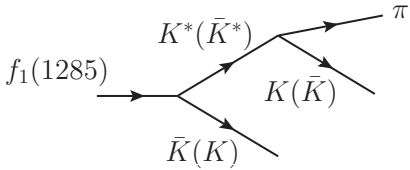}
\caption{\label{Fig4} The diagram of the decay $f_1(1285)\to(K^*\bar
K+\bar K^*K)\to K\bar K\pi$. The four-momenta of =$f_1(1285)$, $K$,
$\bar K$, and  $\pi$ are, respectively, $p_1$, $p_K$, $p_{\bar K}$,
and   $p_\pi$; the four-momenta of the intermediate $K^*$ and  $\bar
K^*$ are $k_1$ and $k_2$, respectively.}\end{figure}

If the meson $f_1(1285)$ decays into $(K^*\bar K+\bar K^*K)\to
 K\bar K\pi$, then, due to the final state interaction of the
$K$ and $\bar K$ mesons, i.e., due to the transitions $K^+K^-\to
f_0(980)\to\pi^+ \pi^-$ and $K^0\bar K^0\to f_0(980)\to\pi^+ \pi^-$,
the isospin-breaking decay  $f_1(1285)\to(K^*\bar K+\bar K^*K)\to(
K^+K^-+K^0\bar K^0)\pi^0$\,$\to$\,$f_0(980)\pi^0$\,$\to$\,$\pi^+
\pi^-\pi^0$ is induced (see Fig.~\ref{Fig3}). This occurs because
the contributions from the
$K^+K^-$ and  $K^0\bar K^0$ pair production are not compensated
entirely. Naturally, the compensation is less pronounced in the
region  $m_{\pi^+ \pi^-}$ ($\sqrt{s}$) between the  $K^+K^-$ and
$K^0\bar K^0$ thresholds. Below we shall obtain the estimate for the
ratio of the branching fractions of the decays
$f_1(1285)\to\pi^+\pi^-\pi^0$ and $f_1(1285)\to K\bar K\pi$, caused
by the mechanisms graphically represented by Figs.~\ref{Fig3} and
\ref{Fig4}, respectively.

The $f_1(1285)\to K^*\bar K$ decay amplitude is determined, in
general, by the two independent effective coupling constants.
However, at the present state of experimental data the general form
of this amplitude is in fact unknown. Hence, for the sake of
definiteness we restrict ourselves with the particular expression
for it (in the spirit of the effective chiral Lagrangian approach
\cite{EGLPR89,EGPR89,Bir96}) of the form
\begin{eqnarray}\label{EqVI-1} V_{f_1K^*\bar K}=g_{f_1K^*\bar K}
F^{(f_1)}_{\,\mu\nu}(F^{(K^*)\mu\nu})^\ast\,, \end{eqnarray} where
$F^{(f_1)}_{\,\mu\nu}=p_{1\mu}\epsilon_{f_1\nu}-p_{1\nu}\epsilon_{f_1
\mu}$, $F^{(K^*)}_{\,\mu\nu}=k_{1\mu}\epsilon_{K^*\nu}-k_{1\nu}
\epsilon_{K^*\mu}$, $\epsilon_{K^*}$ stands for the polarization
four-vector of the $K^*$ meson. Notice that the $K^*$ produced in
the result of such transverse interaction carries the unit spin  off
the mass shell. The  $K^*\to K\pi$ decay amplitude is written as
\begin{eqnarray}\label{EqVI-2}
V_{K^*K\pi}=g_{K^*K\pi}(\epsilon_{K^*},p_\pi-p_K)\,, \end{eqnarray}
where $\,g_{K^{*+}K^+\pi^0}=-g_{\bar K^{*0}\bar K^0\pi^0}$,
$\,g_{K^{*+}K^0\pi^+}=\sqrt{2}g_{{K^{*+}K^+\pi^0}}$. The analogous
expressions are valid for the $f_1(1285)\to\bar K^*K$ and $\bar
K^*\to\bar K\pi$ decays. According to Eqs.~(\ref{EqVI-1})  and
(\ref{EqVI-2}), the product of the vertices in the amplitude of the
diagram shown in Fig.~\ref{Fig3} turns out to be of the third order
in momenta. But two momenta out of three refer to the momenta of
external particles, so that the diagram is convergent (see Appendix
\ref{appB}).

The width of the decay  $f_1(1285)\to(K^*\bar K+\bar K^*K)\to K\bar
K\pi$ (see Fig.~\ref{Fig4}) to all charged modes under assumption of
the isotopic invariance  is written in the form
\begin{eqnarray}\label{EqVI-3} \Gamma_{f_1\to(K^*\bar K+\bar
K^*K)\to K\bar
K\pi}=\frac{g^2_{f_1K^{*+}K^-}g^2_{K^{*+}K^+\pi^0}}{4\,\pi^3\,m^3_{f_1}}
\nonumber\\ \times\int\limits_{(m_K+m_\pi)^2}^{(m_{f_1}-m_K)^2}
dk^2_1 \int\limits_{\tilde{a}_-(k^2_1)}^{\tilde{a}_+(k^2_1)}dk^2_2\
\mathcal{F}(k^2_1,k^2_2)\,,\qquad\end{eqnarray}where
\begin{equation}\label{EqVI-4}
\mathcal{F}(k^2_1,k^2_2)=\frac{-Q^2_1}{|D_{K^*}(k^2_1)|^2}+
\mbox{Re}\,\frac{-(Q_1,Q_2)}{
D_{K^*}(k^2_1)D^*_{K^*}(k^2_2)}\,,\end{equation}
\begin{eqnarray}\label{EqVI-5}
\tilde{a}_\pm(k^2_1)=\frac{1}{2}(m^2_{f_1}+m^2_\pi+2m^2_K-k^2_1)\qquad\ \nonumber\\
+\frac{(m^2_{f_1}-m^2_K)(m^2_K-m^2_\pi)}{2\,k^2_1}
\pm\frac{2m_{f_1}}{\sqrt{k^2_1}}\,\tilde{p}(k^2_1)\tilde{q}(k^2_1)\,,
\end{eqnarray}
\begin{equation} \label{EqVI-5a}
\tilde{p}(k^2_1)=\sqrt{m^4_{f_1}-2m^2_{f_1}(k^2_1+m^2_K)+(k^2_1-m^2_K)^2}\Bigl/(2m_{f_1}),
\end{equation} \begin{equation}\label{EqVI-5b}
\tilde{q}(k^2_1)=\sqrt{k^4_1-2k^2_1(m^2_K+m^2_\pi)+(m^2_K-m^2_\pi)^2}\Bigl/(2\sqrt{k^2_1}),\
\end{equation}
$1/D_{K^*}(k^2_{1(2)})$ stands for the propagator of the  $K^*(\bar K^*)$,
$Q_{1\mu}=(p_1,p_K)p_{\pi\mu}-(p_1,p_\pi)p_{K\mu}$,
$\,Q_{2\mu}=(p_1,p_{\bar K})p_{\pi\mu}-(p_1,p_\pi)p_{\bar K\mu}$;
the functions $Q^2_1$ and $(Q_1,Q_2)$ are given in Appendix \ref{appB}.

The invariant mass of the  $K\pi$ pair, $\sqrt{k^2_1}$, in the decay
$f_1(1285)\to K\bar K\pi$ variates in the interval from  629 to 788
MeV. Since $m_{K^*}\approx895$ MeV and $\Gamma_{K^*}\approx50$
\cite{PDG14}, then, it is easy to convince, the influence of the
width of the virtual intermediate $K^*$ resonance in
Eqs.~(\ref{EqVI-3}) and (\ref{EqVI-4}) turns out to be negligible.
So, we set in what follows that $1/D_{K^*}(k^2_{1(2)})
=1/(m^2_{K^*}-k^2_{1(2)})$, i.e., we neglect the width of the
$K^*(\bar K^*)$ in its propagator.

The numerical integration in Eq.~(\ref{EqVI-3}) gives
\begin{eqnarray}\label{EqVI-6}
\Gamma_{f_1\to(K^*\bar K+\bar K^*K)\to K\bar
K\pi}=\frac{g^2_{f_1K^{*+}K^-}g^2_{K^{*+}K^+\pi^0}}{4\,\pi^3}
\nonumber\\ \times\,0.976\,\times10^{-2}\,\mbox{GeV}^3\,.
\qquad\qquad\end{eqnarray}

\begin{figure} 
\includegraphics[width=6.9cm]{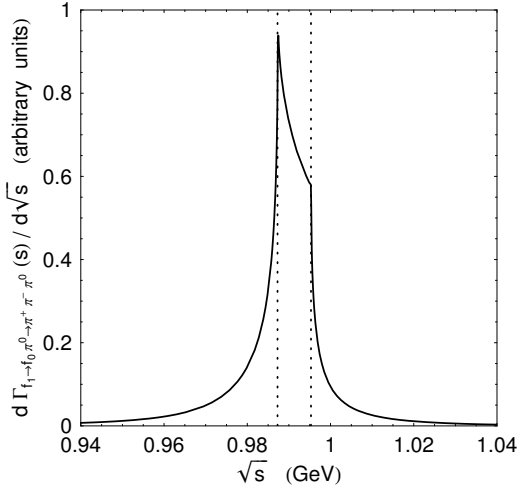}
\caption{\label{Fig5} The $\pi^+\pi^-$ mass spectrum in the decay
$f_1(1285)\to(K^*\bar K+\bar K^*K)\to (K^+K^-+K^0\bar K^0)\pi^0\to
f_0(980)\pi^0\to\pi^+\pi^-\pi^0$.}
\end{figure}

The width of the decay $f_1(1285)\to f_0(980)\pi^0\to\pi^+
\pi^-\pi^0$ in the case of the mechanism shown in Fig.~\ref{Fig3} is
represented in the form
\begin{eqnarray}\label{EqVI-7}\Gamma_{f_1\to f_0\pi^0\to\pi^+\pi^-
\pi^0}\qquad\qquad\qquad  \nonumber \\
=\int\limits_{0.9 \mbox{\scriptsize{ GeV}}}^{1.05\mbox{\scriptsize{
GeV}}}\frac{|G_{f_1
f_0\pi^0}(s)|^2\,p^3(s)}{6\,\pi\,m^2_{f_1}}\,\frac{2s\Gamma_{f_0\to\pi^+\pi^-}(s)}{
\pi|D_{f_0}(s)|^2}\,d\sqrt{s},\ \ \end{eqnarray} where $G_{f_1
f_0\pi^0}(s)$ is the invariant amplitude which determines the
effective vertex of the  $f_1f_0\pi^0$ interaction,
\begin{eqnarray}\label{EqVI-7a}
V_{f_1f_0\pi^0}=G_{f_1f_0\pi^0}(s)(\epsilon_{f_1},p_3-p_2)\,.
\end{eqnarray}
The detailed calculation of  $G_{f_1 f_0\pi^0}(s)$ is give in
Appendix \ref{appB}. The function in the integrand in
Eq.~(\ref{EqVI-7}) gives the mass spectrum of the $\pi^+\pi^-$ pair,
$d\Gamma_{f_1\to f_0\pi^0\to\pi^+ \pi^-\pi^0 }(s)/d\sqrt{s}$. Its
sharp enhancement in the region of the $K\bar K$ thresholds (see
Fig.~\ref{Fig5}) is determined by the corresponding behavior of the
amplitude $G_{f_1f_0\pi^0}(s)$.
Let us turn attention to the fact that the shape of the $\pi^+\pi^-$
spectrum in Fig.~\ref{Fig5} practically coincides with the
corresponding spectrum shown in Fig.~\ref{Fig1}, caused by the
$a^0_0(980)-f_0(980)$ mixing. The integration in Eq.~(\ref{EqVI-7})
gives
\begin{eqnarray}\label{EqVI-8}
\Gamma_{f_1\to f_0\pi^0\to\pi^+\pi^- \pi^0}=\frac{g^2_{
f_1K^{*+}K^-}g^2_{K^{*+}K^+\pi^0}}{4\,\pi^3} \nonumber\\
\times\,0.277\,\times10^{-4}\,\mbox{GeV}^3\,.\qquad\qquad\end{eqnarray}
Comparing Eq.~(\ref{EqVI-8}) with Eq.~(\ref{EqVI-6}) we obtain
\begin{eqnarray}\label{EqVI-9}
\frac{\Gamma_{f_1\to f_0\pi^0\to\pi^+\pi^-\pi^0}}{\Gamma_{
f_1\to(K^*\bar K+\bar K^*K)\to K\bar K\pi}}\qquad\quad \nonumber\\
=\frac{BR(f_1\to f_0\pi^0\to\pi^+\pi^-\pi^0)} {BR(f_1\to(K^*\bar
K+\bar K^*K)\to K\bar K\pi)}\nonumber\\
=0.284\,\times10^{-2}\,.\qquad\quad\quad
\end{eqnarray}

Assuming that the whole decay branching  $BR(f_1\to K\bar
K\pi)=(9.0\pm0.4)\%$ \cite{PDG14} results from the  $f_1\to(K^*\bar
K+\bar K^*K)\to K\bar K\pi$ decay mode, the following estimate for
$BR(f_1\to f_0\pi^0\to\pi^+\pi^-\pi^0)$ follows from
Eq.~(\ref{EqVI-9}):
\begin{eqnarray}\label{EqVI-10}
BR(f_1\to f_0\pi^0\to\pi^+\pi^-\pi^0)\approx 0.255\,\times10^{-3}\,.
\end{eqnarray} This value is approximately 12 times lower than
the central experimental value in Eq.~(\ref{EqII-2}) obtained by
VES. If, in addition, one takes into account the relations
(\ref{EqIV-11}) and (\ref{EqIV-12}), then the estimate
Eq.~(\ref{EqVI-10}) should be further divided by approximately 1.8.
So, the $f_1(1285)\to(K^*\bar K+ \bar K^*K)\to (K^+K^-+K^0\bar
K^0)\pi^0\to f_0(980)\pi^0\to\pi^+ \pi^-\pi^0$ transition mechanism
alone is also insufficient to understand the experimental data.


\section{\boldmath DECAY $f_1(1285)\to(K^*_0\bar K+\bar K^*_0
K)\to(K^+K^-+K^0\bar K^0)\pi^0\to f_0(980)\pi^0\to\pi^+\pi^-\pi^0$}
\label{sec7}
~

Let us try to reveal the possible role of the decay mechanism
$f_1(12 85)\to(K^*_0\bar K+\bar K^*_0 K)\to(K^+K^-+K^0\bar
K^0)\pi^0\to f_0 (980)\pi^0\to\pi^+\pi^-\pi^0$ with the
participation of the scalar meson $K^*_0$. The variant with the
$K^*_0(800)$ resonance (or $\kappa$) [37] should be rejected. The
fact is that for the $\kappa$ resonance with the mass $m_\kappa$
which is approximately equal or less than  800 MeV and the width
$\Gamma_\kappa\approx\,$(400--550) MeV \cite{PDG14}, the shapes of
the $K\pi$ and  $K\bar K$ spectra in the decay
$f_1(1285)\to(\kappa\bar K+\bar\kappa K)\to K\bar K\pi$ are
literally opposite to those observed in the experiment
\cite{Bit84,Arm84,Arm87}. According to the data on the $f_1(1285)\to
K\bar K\pi$ decay \cite{PDG14,Bit84,Arm84,Arm87}, there is
considerable enhancement in the $K\bar K$ spectrum near the $K\bar
K$ threshold, while in the $K\pi$ mass spectrum one observes the
large enhancement near the upper border of the spectrum, i.e., near
$m_{f_1}-m_K\approx788$ MeV. Such a picture agrees well with the
$f_1(1285)\to a_0(980)\pi\to K\bar K\pi$ decay mechanism and does
not contradict to the mechanism $f_1(1285)\to(K^*\bar K+\bar
K^*K)\to K\bar K\pi$. On the contrary, the $f_1(1285)\to(\kappa\bar
K+\bar \kappa K)\to K\bar K\pi$ mechanism results in the sharp
enhancement near the upper border of the $K\bar K$ spectrum, i.e.,
near $m_{f_1}-m_\pi\approx1147$ MeV, and to the enhancement of the
$K\pi$ spectrum close to its threshold. Clearly, such a mechanism
cannot be responsible for  a sizable portion of the decay
$f_1(1285)\to K\bar K\pi$. Also, we cannot point to some special
enhancement of the decay $f_1(1285)\to f_0(980)\pi^0\to\pi^+
\pi^-\pi^0$ due to this mechanism.

Increasing the mass of the  $K^*_0$ resonance (pushing it from the
$K\pi $ threshold, $m_K+m_\pi\approx0.629$ GeV) makes the
disagreement with the data on the  $K\bar K$ and $K\pi$ mass spectra
less pronounced. The resonance  $K^*_0(1430)$ with the mass
$m_{K^*_0}\approx1425$ MeV and the width
$\Gamma_{K^*_0}\approx270$MeV \cite{PDG14} could be considered as
the candidate responsible  for the decay  $f_1(1285)\to(K^*_0 \bar
K+\bar K^*_0 K)\to(K^+K^-+K^0\bar K^0)\pi^0\to f_0(980)\pi^0\to
\pi^+\pi^-\pi^0$.

Along the lines similar to Sec.~\ref{sec6}, first  let us calculate
the width of the decay $f_1(1285)\to( K^*_0 \bar K+\bar K^*_0 K)\to
K\bar K\pi$ to all charge modes  [see Fig.\ref{Fig4}, where the
resonance $K_0^*(\bar K_0^*)$ should be inserted instead of
$K^*(\bar K^*)$]. The amplitude of the  $f_1(1285)\to K^*_0\bar K$
transition looks as  $V_{f_1K_0^*\bar K}=g_{f_1K_0^*\bar
K}(\epsilon_{f_1},p_{\bar K}-p_{K_0^*})$ [analogously for the
$f_1(1285)\to\bar K^*_0K$ one]. One has, assuming the isotopic
symmetry,
\begin{eqnarray}\label{EqVII-1} \Gamma_{f_1\to(K_0^*\bar K+\bar
K_0^*K)\to K\bar
K\pi}=\frac{g^2_{f_1K_0^{*+}K^-}g^2_{K_0^{*+}K^+\pi^0}}{
16\,\pi^3\,m^3_{f_1}} \nonumber\\
\times\int\limits_{(m_K+m_\pi)^2}^{(m_{f_1}-m_K)^2} dk^2_1
\int\limits_{\tilde{a}_-(k^2_1)}^{\tilde{a}_+(k^2_1)}dk^2_2\
\widetilde{\mathcal{F}}(k^2_1,k^2_2)\,,\qquad\end{eqnarray} where
\begin{eqnarray}\label{EqVII-2}
\widetilde{\mathcal{F}}(k^2_1,k^2_2)=\frac{|\tilde{p}(k^2_1)|^2}{|D_{K_0^*}
(k^2_1)|^2}+ \mbox{Re}\,\frac{\tilde{p}(k^2_1)\tilde{p}(k^2_2)\cos
\tilde{\theta}}{ D_{K_0^*}(k^2_1)D^*_{K_0^*}(k^2_2)}\,,
\end{eqnarray}
\begin{eqnarray}\label{EqVII-3}
\tilde{p}(k^2_1)\tilde{p}(k^2_2)\cos\tilde{\theta}=\frac{1}{2}
(k^2_1+k^2_2-m^2_{f_1}-m^2_\pi) \quad\nonumber \\
+\frac{(m^2_{f_1}+m^2_K-k^2_1)
(m^2_{f_1}+m^2_K-k^2_2)}{4m^2_{f_1}}\,,\qquad \end{eqnarray} and
$1/D_{K_0^*}(k^2_{1(2)})$ is the propagator of the $K_0^*(\bar
K_0^*)$. In what follows we set
$1/D_{K_0^*}(k^2_{1(2)})=1/(m^2_{K_0^*}-k^2_{1(2)})$, i.e., we
neglect the width of the $K_0^*(\bar K_0^*)$ resonance in its
propagator. This is a good approximation for the $f_1(1285)\to(K^*_0
\bar K+\bar K^*_0 K)\to K\bar K\pi$ decay that considerably
simplifies the calculations. The numerical integration in
Eq.~(\ref{EqVII-1}) gives
\begin{eqnarray}\label{EqVII-4}
\Gamma_{f_1\to(K_0^*\bar K+\bar K_0^*K)\to K\bar
K\pi}=\frac{g^2_{f_1K_0^{*+}K^-}g^2_{K_0^{*+}K^+\pi^0}}{16\,\pi^3}
\nonumber\\ \times\,0.971\,\times10^{-4}\,\mbox{GeV}^{-1}\,.
\qquad\qquad\end{eqnarray}

Notice that the strong destructive interference occurs between the
$K^*_0\bar K$ and  $\bar K^*_0K$ intermediate state contributions in
the decay  $f_1(1285)\to(K^*_0 \bar K+\bar K^*_0K)\to K\bar K\pi$.
Namely, the second interfering term in Eq.~(\ref{EqVII-2}) turns out
to be large in magnitude and negative in practically the entire
physical region  of the variables  $k^2_1$ and $k^2_2$. As a result,
the interference reduces the result obtained without interference
taking into account by approximately  74\%.

The  $f_1(1285)\to f_0(980)\pi^0\to\pi^+\pi^-\pi^0$ transition width
for the mechanism  $f_1(1285)\to(K^*_0\bar K+\bar K^*_0
K)\to(K^+K^-+K^0\bar K^0)\pi^0\to f_0(980)\pi^0\to\pi^+\pi^- \pi^0$
[see Fig.~\ref{Fig3} in which $K_0^*(\bar K_0^*)$ should be
substituted instead of $K^*(\bar K^*)$] can be represented in the
form
\begin{eqnarray}\label{EqVII-5}\Gamma_{f_1\to f_0\pi^0\to\pi^+\pi^-
\pi^0}\qquad\qquad\qquad  \nonumber \\
=\int\limits_{0.9 \mbox{\scriptsize{ GeV}}}^{1.05\mbox{\scriptsize{
GeV}}}\frac{|\widetilde{G}_{f_1 f_0\pi^0}(s)|^2\,p^3(s)}{6\,
\pi\,m^2_{f_1}}\,\frac{2s\Gamma_{f_0\to\pi^+\pi^-}(s)}{
\pi|D_{f_0}(s)|^2}\,d\sqrt{s},\end{eqnarray} where
$\widetilde{G}_{f_1 f_0\pi^0}(s)$ is the invariant amplitude that
determines the effective vertex of the $f_1f_0\pi^0$ interaction,
\begin{eqnarray}\label{EqVII-5a}
\widetilde{V}_{f_1f_0\pi^0}=\widetilde{G}_{f_1f_0\pi^0}(s)(\epsilon_{
f_1},p_3-p_2)\,.\end{eqnarray} The evaluations of
$\widetilde{G}_{f_1f_0 \pi^0}(s)$ and  $\Gamma_{f_1\to
f_0\pi^0\to\pi^+ \pi^-\pi^0}$ are analogous to those made in
Sec.~\ref{sec6} and Appendix \ref{appB}. We will not dwell on them
here. We restrict ourselves only  by  pointing  out that the
$\pi^+\pi^-$ mass spectrum in the decay $f_1(1285)\to(K^*_0\bar
K+\bar K^*_0 K)\to(K^+K^-+K^0\bar K^0)\pi^0\to
f_0(980)\pi^0\to\pi^+\pi^- \pi^0$ looks similar to the $\pi^+\pi^- $
mass spectra in Figs.~\ref{Fig1} and \ref{Fig5} and cite the final
result of the $\Gamma_{f_1\to f_0\pi^0\to\pi^+ \pi^-\pi^0}$
evaluation:
\begin{eqnarray}\label{EqVII-6} \Gamma_{f_1\to f_0\pi^0\to\pi^+
\pi^- \pi^0}=\frac{g^2_{f_1K_0^{*+}K^-}g^2_{K_0^{*+}K^+\pi^0}
}{16\,\pi^3} \nonumber\\ \times\,0.263\,\times10^{-6}\,
\mbox{GeV}^{-1}\,.\qquad\qquad\end{eqnarray}

The comparison of  (\ref{EqVII-6}) with (\ref{EqVII-4}) gives
\begin{eqnarray}\label{EqVII-7}
\frac{\Gamma_{f_1\to f_0\pi^0\to\pi^+\pi^-\pi^0}}{\Gamma_{
f_1\to(K_0^*\bar K+\bar K_0^*K)\to K\bar K\pi}}\qquad\quad \nonumber\\
=\frac{BR(f_1\to f_0\pi^0\to\pi^+\pi^-\pi^0)}{BR(f_1\to(K_0^*\bar
K+\bar K_0^*K)\to K\bar K\pi)}\nonumber\\
=0.271\,\times10^{-2}\,.\qquad\quad\quad
\end{eqnarray}
Since the estimate  (\ref{EqVII-7}) practically coincides with
Eq.~(\ref{EqVI-9}), then the statements made about  $BR(f_1\to
f_0\pi^0\to\pi^+ \pi^- \pi^0)$ after Eq.~(\ref{EqVI-9}) are valid in
the present case, too. So, the mechanism of the
$f_1(1285)\to(K_0^*\bar K+ \bar K_0^*K)\to(K^+K^-+K^0\bar
K^0)\pi^0\to f_0(980)\pi^0\to\pi^+ \pi^-\pi^0$ transition cannot by
itself explain the experimental data.

\section{ General approach to description of the $K\bar K$ loop breaking of isotopic invariance}
\label{sec8} ~

\begin{figure} [!ht] 
\includegraphics[width=6.0cm]{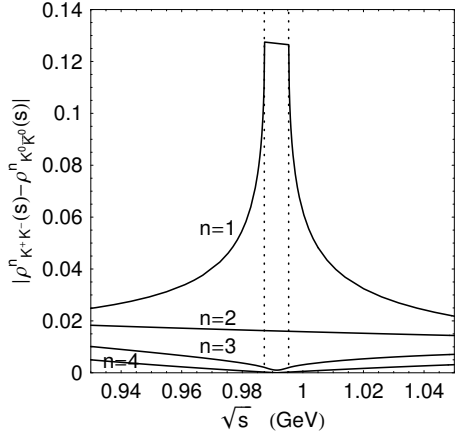}
\caption{\label{Fig7} The functions $|\rho^n_{K^+K^-}(s)-\rho^n_{K^0\bar
K^0}(s)|$ for $n=1,2,3,$ and 4 near the $K\bar K$ thresholds.}
\end{figure}

Let us write the  $\pi^+\pi^-$ mass spectrum in the decay  $f_1(1285)\to
f_0(980)\pi^0\to\pi^+\pi^-\pi^0$ in the form
\begin{eqnarray}\label{EqVIII-1}\frac{d\Gamma_{f_1\to f_0\pi^0\to\pi^+
\pi^-\pi^0}(s)}{d\sqrt{s}}\qquad\qquad \nonumber \\
=\frac{1}{16\pi}|\mathcal{M}_{f_1\to
f_0\pi^0}(s)|^2\,p^3(s)\,\frac{2s\Gamma_{f_0\to\pi^+\pi^-}(s)}{
\pi|D_{f_0}(s)|^2}\,.\end{eqnarray} The isospin breaking amplitude
$\mathcal{M}_{f_1\to f_0\pi^0}(s)$ can be expanded near the  $K\bar K$
threshold into the series in $\rho_{K\bar K}(s)=\sqrt{1-4 m^2_K/s}$:
\begin{eqnarray}\label{EqVIII-2}
\mathcal{M}_{f_1\to f_0\pi^0}(s)=g_{f_0K^+K^-}\left\{A(s)\right. \
\qquad\nonumber\\ \times i[\rho_{K^+K^-}(s)-\rho_{K^0\bar K^0}(s)]
+B(s)[\rho^2_{K^+K^-}(s) \nonumber \\ -\rho^2_{K^0 \bar K^0}(s)]
+O[\rho^3_{K^+K^-}(s)-\rho^3_{K^0\bar K^0}(s)]+\cdot\cdot\cdot
\left.\right\}.
\end{eqnarray}
The character of the behavior of the functions
$|\rho^n_{K^+K^-}(s)-\rho^n_{K^0\bar K^0}(s)|$ near the  $K\bar K$
threshold is shown in Fig.~\ref{Fig7}.

Let us restrict ourselves in Eq.~(\ref{EqVIII-2}) by the dominant
term proportional to $i[\rho_{K^+K^-}(s)-\rho_{K^0\bar K^0}(s)]$\,
[1], i.e., let us set
\begin{eqnarray}\label{EqVIII-3}\mathcal{M}_{f_1\to f_0
\pi^0}(s)=g_{f_0K^+K^-}\ \ \ \nonumber\\
\times A(s)i[\rho_{K^+K^-}(s)-\rho_{K^0\bar K^0}(s)].
\end{eqnarray} The amplitude $A(s)$ contains the information about
all possible mechanisms of production of the  $K\bar K$ system with
isospin  $I=1$ in $S$ wave  in the process $f_1(1285)\to K\bar
K\pi$. From the data on the decay $f_1(1285)\to
f_0(980)\pi^0\to\pi^+\pi^-\pi^0$\, one can extract the information
about  $|A(s)|^2$ in the region above the $K^0\bar K^0$ threshold,
between the $K^+K^-$ and  $K^0\bar K^0$ thresholds, and below the
$K^+K^- $ threshold. A simplest variant of the description of the
data on $d\Gamma_{f_1\to f_0\pi^0\to \pi^+\pi^-\pi^0}(s)/d\sqrt{s}$
with the help of Eqs.~(\ref{EqVIII-1}) and (\ref{EqVIII-3}) can be
realized by setting  $|A(s)|^2$ to be constant, for instance, upon
setting  $|A(s)|^2=|A(4m^2_{K^+})|^2$). Then resulting from this fit
of the $\pi^+\pi^-$ mass spectrum will be determination of this
constant.

The information about $|A(s)|^2$ at $\sqrt{s}>2m_K$ can be obtained
from the data on the $K\bar K$ mass spectra measured in the decays
$f_1(1285)\to K\bar K\pi$. Unfortunately, the data on these spectra
are poor as yet [53-55]. However, possessing the high statistics and
good resolution in the invariant mass of the  $K\bar K$
($\sqrt{s}$), the simple scheme of obtaining the information about
$|A(s)|^2$ at $\sqrt{s}$ above the  $K^+K^-$, or $K^0\bar K^0$, or
$K^\pm K^0_S$ thresholds could be consisted in the following.

The $K\bar K$ system, due to the essential restriction of the
admissible phase space in the decay  $f_1(1285)\to K\bar K\pi$
($2m_K< \sqrt{s}<2m_K+150$\,MeV), should be produced predominantly
in $S$ wave. Then, for instance, the $K^+K^-$ spectrum in the decay
$f_1(1285)\to K^+K^-\pi^0$ can be represented in the form
\begin{eqnarray}\label{EqVIII-4} \frac{d\Gamma_{f_1\to
K^+K^-\pi^0}}{d\sqrt{s}} =\frac{2\,\sqrt{s}}{\pi}\,\rho_{K^+K^-}(s)
\,p^3(s)\,|A(s)|^2\,.\end{eqnarray} Fitting the data on
$d\Gamma_{f_1\to K^+K^-\pi^0}/d\sqrt{s}$, one can construct the
function $|A(s)|^2$. Using its value at the $K^+K^-$ threshold,  $|A
(4m^2_{K^+})|^2$ (which, for granted, corresponds to the
contribution of the $S$ wave) and Eqs.~(\ref{EqVIII-1}) and
(\ref{EqVIII-3}), one can obtain the estimate for the quantity
\begin{eqnarray}\label{EqVIII-5} \Gamma_{f_1\to f_0\pi^0\to
\pi^+\pi^-\pi^0}=\int\frac{d\Gamma_{f_1\to f_0\pi^0\to\pi^+
\pi^-\pi^0}(s)}{d\sqrt{s}}\,d\sqrt{s}\nonumber\\ =\int|A
(4m^2_{K^+})|^2|\rho_{K^+K^-}(s)-\rho_{K^0\bar K^0}(s)|^2\qquad
\nonumber\\ \times p^3(s)\,\frac{g^2_{f_0K^+K^-}}{16\pi} \,
\frac{2s\Gamma_{f_0\to\pi^+ \pi^-}(s)}{\pi|D_{f_0}(s)|^2}\,\sqrt{s}
\,.\qquad\end{eqnarray} Its comparison with the data on the decay
$f_1(1285) \to\pi^+\pi^-\pi^0$ permits one to verify their
consistence with the data on the decay  $f_1(1285)\to K\bar K\pi$
and with the idea of the breaking of isotopic invariance caused by
the mass difference of  $K^+$ and  $K^0$ mesons.

Upon using the coupling constants found by us, the integration in
Eq.~(\ref{EqVIII-5}) over the interval from 0.9 to 1.05 GeV gives
\begin{eqnarray}\label{EqVIII-6}\Gamma_{f_1\to f_0\pi^0\to
\pi^+\pi^-\pi^0}=|A (4m^2_{K^+})|^2\,2.59\times10^{-6}
\,\mbox{GeV}^5.\ \ \end{eqnarray}

The proposed approach is applicable to the estimates of other decays
of similar sort.

If the isospin-violating amplitude contains in the physical region
of kinematic variables (in the region of the $K\bar K$ thresholds)
the logarithmic (triangle) singularities, as in the case of the
$\eta(1405)\to(K^*\bar K+\bar K^*K)\to(K^+K^-+K^0\bar K^0)\pi^0\to
f_0(980)\pi^0\to\pi^+\pi^-\pi^0$ decay, then its structure near the
$K\bar K$ thresholds becomes more sophisticated and the consistency
condition of the type of Eq. (\ref{EqVIII-5}) cannot be obtained.

\section{SOME COMMENTS ABOUT ESTIMATES}
\label{sec9} ~

The effect under consideration is caused by the $K$ meson mass
difference and manifests itself in the vicinity of the $K^+K^-$ and
$K^0\bar K^0$ thresholds, where kaons are near their mass shells.
All resonance contributions taken into account by us in the
intermediate states of the tree diagrams and in the imaginary parts
of the triangle loop diagrams appear also near the mass shells, that
is, at $s\approx m^2_{res}$, etc. It means that the vertex form
factors, usually suppressing the hadronic amplitudes, do not play an
essential role in the present case. As for the real parts of the
triangle loop diagrams, the insertion of the form factor for
obtaining their numerical estimate would have some meaning in the
case of the divergent diagrams. In our case, the triangle diagrams
with the charged and neutral intermediate kaon states are convergent
separately, hence their estimates are possible without introduction
of any phenomenological form factors. Moreover, the result of
compensation of the charged and neutral intermediate states in the
channels $f_1(1285)\to(K^*\bar K+\bar K^*K)\to(K^+ K^- +K^0\bar K^0)
\pi^0\to f_0(980)\pi^0\to \pi^+\pi^-\pi^0$ and $f_1(1285)\to(K^*_0
\bar K+\bar K^*_0K)\to(K^+K^-+K^0\bar K^0)\pi^0\to f_0(980)\to\pi^+
\pi^-\pi^0$ (i.e., the shape of the basic contribution near the
$K\bar K$ thresholds) turns out to be practically insensitive to the
form factor behavior off the mass shell.

It should be also emphasized  that the convergence or divergence of
the triangle diagrams as well as of the $K\bar K$ loops in the case
of the $a^0_0(980)\to(K^+K^-+K^0\bar K^0)\to f_0(980)$ transition is
not related with the effect under discussion. The sum of the
subtraction constants for the contributions of the charged and
neutral intermediate states in the dispersion representation for the
isospin breaking amplitude should have the natural order of
smallness $\sim(m_{K^0}-m_{K^+})$, and it cannot be responsible for
the enhancement of the symmetry violation in the vicinity of the
$K^+K^-$ and $K^0\bar K^0$ thresholds neither in the magnitude nor
in the shape.

Let us call attention to the fact that the estimates obtained for
the ratios  (4.10), (5.3), (6.12), and (7.8), which characterize the
isospin breaking for different mechanisms, do not depend on the
magnitudes of the coupling constants $g_{f_1a^0_0\pi^0}$,
$g_{f_1K^+K^-\pi^0}$, $g_{f_1K^{*+}K^-}$, and $g_{f_1K^{*+}_0K^-}$,
respectively. By themselves, these constants are either ill defined
or simply unknown. Hence, in order to combine meaningfully the
different theoretical mechanisms of the decay $f_1(1285)\to
f_0(980)\pi^0\to\pi^+\pi^-\pi^0$, the considerably improved quality
of the data on the main decay channels $f_1(1285)\to\eta \pi\pi$ and
$f_1(1285)\to K\bar K\pi$ is necessary. The partial wave analysis of
the three-particle events is required for the clarification of the
relative role of the specific mechanisms in the above channels. In
this route, we would persuade the experimenters to measure in the
first place the decays $f_1(1285)\to\pi^+\pi^-\pi^0$ and
$f_1(1285)\to K^+ K^-\pi^0$ simultaneously (at the same facility and
in the same experiment) and to obtain the $\pi^+\pi^-$- and $K^+
K^-$ mass spectra. As it is explained in Sec. VIII, this would give
the possibility of checking the consistency of the data on the
$\pi^+\pi^-$- and $K^+ K^-$ mass spectra before the detailed
partial-wave analysis.

\section{CONCLUSION AND OUTLOOK}
\label{sec10} ~

The phenomenon of the $a^0_0(980)-f_0(980)$ mixing  \cite{ADS79}
gave the impetus to the doing experiments on reactions  (a)--(f)
which were made by the collaborations VES \cite{Do08,Do11} and
BESIII \cite{Ab1,Ab2,Ab3}. In the present work, we show the
principal possibility of the estimate of coupling constants of the
$a_0(980)$ and  $f_0(980)$ resonances using the BESIII data
\cite{Ab1} on the  $a_0(980)-f_0(980)$ mixing. Notice that the
relations among the couplings found in Sec.~\ref{estimates} agree
well with the predictions of the  $q^2\bar q^2$ model. Interesting
for physics and promising problem is the task of making more precise
the BESIII data \cite{Ab1} on reactions (b) and (c) [see
Eqs.~(\ref{EqIII-1}) and (\ref{EqIII-2})].

We have analyzed in detail four possible mechanisms for the
isospin-breaking decay  $f_1(1285)\to\pi^+\pi^-\pi^0$:
\begin{enumerate}
\item $f_1(1285)\to a_0(980)\pi^0\to (K^+K^-+K^0\bar
K^0)\pi^0\to f_0(980)\pi^0\to\pi^+\pi^-\pi^0$,
\item $f_1(1285)\to(K^+K^-+K^0\bar K^0)\pi^0\to f_0
(980)\pi^0\to\pi^+\pi^-\pi^0$,
\item $f_1(1285)\to(K^*\bar K+\bar
K^*K)\to(K^+K^-+K^0\bar K^0)\pi^0\to f_0(980)\pi^0\to
\pi^+\pi^-\pi^0$,
\item $f_1(1285)\to(K^*_0\bar K+\bar K^*_0
K)\to(K^+K^-+K^0\bar K^0)\pi^0\to f_0(980)\pi^0\to\pi^+
 \pi^-\pi^0$.\end{enumerate}
Our conclusions from the estimates are the following. The
experimental data are difficult to explain by the single specific
mechanism from those listed above. On the other hand, these
mechanisms are united by the fact that for each of them the
$\pi^+\pi^- $ mass spectrum in the decay
$f_1(1285)\to\pi^+\pi^-\pi^0$ turns out to be located between the
$K^+K^-$ and $K^0\bar K^0$ thresholds in view of the $K\bar K$ loop
mechanism of the isospin breaking. It is clear that the considered
mechanisms of the decay $f_1(1285)\to\pi^+\pi^-\pi^0$ underlie the
observable isospin breaking phenomenon. It is apparent also that
considerable experimental efforts are yet required to eliminate the
uncertainties in the available data [for example, it is desirable to
measure the various decay modes of $f_1(1285)$ simultaneously at the
same experimental setup].

Taking the decay  $f_1(1285)\to f_0(980)\pi^0\to\pi^+\pi^-\pi^0$ as
an example we have discussed also the general approach to the
description of the $K\bar K$ loop mechanism of the breaking of
isotopic invariance in the absence of logarithmic singularities.

Since the existing data on the $f_1(1285)\to\pi^+\pi^-\pi^0$ decay
probability have a rather large spread [see Eqs.
(\ref{EqII-3})--(\ref{EqII-5})], the task of making them to be
precise is an extremely interesting and important problem.

Among the numerous production reactions of the $f_1(1285)$ resonance
we want to call attention to the reaction of the $f_1(1285)$
production in the central region via the two-pomeron exchange, and
to the possibility of the study in this reaction the decay
$f_1(1285)\to\pi^+\pi^-\pi^0$,
$$pp\to p_ff_1(1285)p_s\to p_f(\pi^+\pi^-\pi^0)p_s\,.$$
It is interesting also to study the related reaction $pp\to
p_ff_1(1420)p_s\to p_f(\pi^+\pi^-\pi^0)p_s\,.$

$$\mbox{\small \bf ACKNOWLEDGMENTS}$$

The present work is partially supported by the Russian Foundation
for Basic Research Grant No. 16-02-00065 and the Presidium of the
Russian Academy of Sciences project No. 0314-2015-0011.


\appendix
\section{Polarization operators} \label{appA} ~

The polarization operator  $\Pi^{ab}_{r}(s)$ [see
Eqs.~(\ref{EqIII-8}) and (\ref{EqIII-9})] can be written as
\begin{eqnarray}\label{Eq1A}\Pi^{ab}_{r}(s)=g^2_{r
ab}\,\widetilde{B}_0(s;m_a,m_b)\,.\end{eqnarray} The function
$\widetilde{B}_0(s;m_a,m_b)$ at $s>m_{ab}^{(+)\,2}$ looks as
\begin{eqnarray}\label{Eq2A}\widetilde{B}_0(s;m_a,m_b)=\frac{1}{16\pi}
\left[\frac{m_{ab}^{(+)}m_{ab}^{(-)}}{\pi
s}\ln\frac{m_b}{m_a}+\rho_{ab}(s)\right.\ \nonumber\\
\left.\times\left(i-\frac{1}{\pi}\,\ln\frac{\sqrt{s-m_{ab}^{(-)
\,2}}+\sqrt{s-m_{ab}^{(+)\,2}}}{\sqrt{s-m_{ab}^{(-)\,2}}-\sqrt{s
-m_{ab}^{(+)\,2}}}\right)\right],\end{eqnarray} where
$\rho_{ab}(s)$\,=\,$\sqrt{s-m_{ab}^{(+)\,2}}
\,\sqrt{s-m_{ab}^{(-)\,2}}\Bigl/s$, $m_{ab}^{(\pm)}=m_a\pm m_b$ and
$m_a\geq m_b$. At $m_{ab}^{(-)\,2}<s<m_{ab}^{(+)\,2}$
\begin{eqnarray}\label{Eq3A}\widetilde{B}_0(s;m_a,m_b)=\frac{1}{16\pi}
\left[\frac{m_{ab}^{(+)}m_{ab}^{(-)}}{\pi
s}\ln\frac{m_b}{m_a}\right.\nonumber\\
\left.-\rho_{ab}(s)\left(1-\frac{2}{\pi}\arctan\frac{\sqrt{
m_{ab}^{(+)\,2}-s}}{\sqrt{s-m_{ab}^{(-)\,2}}}\right)\right],\end{eqnarray}
where  $\rho_{ab}(s)$\,=\,$\sqrt{m_{ab}^{(+)\,2}-s}
\,\sqrt{s-m_{ab}^{(-)\,2}}\Bigl/s$. At $s\leq m_{ab}^{(-)\,2}$
\begin{eqnarray}\label{Eq4A}\widetilde{B}_0(s;m_a,m_b)=\frac{1}{16\pi}
\left[\frac{m_{ab}^{(+)}m_{ab}^{(-)}}{\pi
s}\ln\frac{m_b}{m_a}\right.\ \ \nonumber\\
\left.+\rho_{ab}(s)\frac{1}{\pi}\,\ln\frac{
\sqrt{m_{ab}^{(+)\,2}-s}+\sqrt{m_{ab}^{(-)\,2}-s}}
{\sqrt{m_{ab}^{(+)\,2}-s}-\sqrt{m_{ab}^{(-)\,2}-s}}\right],\end{eqnarray}
where $\rho_{ab}(s)$\,=\,$\sqrt{m_{ab}^{(+)\,2}-s}\,\sqrt{
m_{ab}^{(-)\,2}-s}\Bigl/s$.


\section{Triangle diagram} \label{appB} ~

The functions  $Q^2_1$ and $(Q_1,Q_2)$ in Eq.~(\ref{EqVI-4}) look as
follows:
\begin{eqnarray}\label{Eq1B} &
Q^2_1=\frac{1}{4}\Bigl[m^2_\pi(m^2_{f_1}+m^2_K-k^2_2)^2+m^2_K
(k^2_1+k^2_2 \nonumber\\ & -2m^2_K)^2-(m^2_{f_1}+m^2_K-k^2_2)
(k^2_1+k^2_2-2m^2_K) \nonumber\\ &
\times(k^2_1-m^2_\pi-m^2_K)\Bigl]\,,\end{eqnarray}
\begin{eqnarray}\label{Eq2B}
(Q_1,Q_2)=\frac{1}{8}\Bigl[2m^2_\pi(m^2_{f_1}+m^2_K-k^2_2)
(m^2_{f_1}+m^2_K \nonumber\\ -k^2_1)
+(k^2_1+k^2_2-2m^2_K)^2(m^2_{f_1}+m^2_\pi-k^2_1-k^2_2)\nonumber\\
-(m^2_{f_1}+m^2_K-k^2_2)(k^2_1+k^2_2-2m^2_K) \nonumber\\
\times(k^2_2-m^2_\pi-m^2_K)-(m^2_{f_1}+m^2_K-k^2_1) \nonumber\\
\times(k^2_1+k^2_2-2m^2_K)(k^2_1-m^2_\pi-m^2_K)\Bigl].
\end{eqnarray}

The invariant amplitude $G_{f_1 f_0\pi^0}(s)$ introduced in
Eqs.~(\ref{EqVI-7}) and (\ref{EqVI-7a}) is represented as the sum of
the amplitudes corresponding to the charged  $(c)$ and neutral $(n)$
intermediate states in the kaon triangle loop in Fig.~\ref{Fig3},
\begin{eqnarray} \label{Eq3B} G_{f_1 f_0\pi^0}(s)=
G^{(c)}_{f_1 f_0\pi^0}(s)+G^{(n)}_{f_1 f_0\pi^0}(s)\,.\end{eqnarray}
In the approximation of the isotopic invariance for coupling
constants of the $f_1(1285)$, $K^*$, and $f_0(980)$ resonances and
at $m_{K^{*+}}=m_{K^{*0}}$, the amplitude   $G^{(c)}_{f_1
f_0\pi^0}(s)$ and $G^{(n)}_{f_1 f_0\pi^0}(s)$ differ by only the
overall  sign and by the masses of the  $K^+(K^-)$ and  $K^0(\bar
K^0)$ mesons.

Using Eqs.~(\ref{EqVI-1}) and (\ref{EqVI-2}), let us write the
amplitude of the triangle diagram in Fig.~\ref{Fig3} for the charged
intermediate kaon states in the following way:
\begin{eqnarray} \label{Eq4B} V^{(c)}_{f_1
f_0\pi^0}=\bar{g}_{f_1f_0\pi^0}\,C^{(c)}_\mu\left[(
\epsilon_{f_1},p_3)p_{1\mu}-(p_1,p_3)\epsilon_{f_1\mu}\right],
\end{eqnarray} where $\bar{g}_{f_1f_0\pi^0}=2(2g_{f_1
K^{*+}K^-}2g_{K^{*+}K^+\pi^0}g_{f_0K^{+}K^-})$ and
\begin{eqnarray} \label{Eq5B} C^{(c)}_\mu=\frac{i}{(2\pi)^4}\int
\frac{k_\mu\,d^4k}{(k^2-m^2_{K^*})((p_1-k)^2-m^2_{K^+})} \nonumber\\
\times\,\frac{1}{((k-p_3 )^2-m^2_{K^-})}.\ \ \end{eqnarray}
Expanding the amplitude $C^{(c)}_\mu$ in the momenta of external
particles, $C^{(c)}_\mu=p_{1\mu}C^{(c)}_{11}+p_{ 3\mu }C^{(c)}_{12}$
\cite{PaV79}, we can rewrite Eq.~(\ref{Eq4B}) in the form
\begin{eqnarray} \label{Eq6B}V^{(c)}_{f_1 f_0\pi^0}=G^{(c)}_{f_1
f_0\pi^0}(s)\,(\epsilon_{f_1}, p_3-p_2)\end{eqnarray} and determine
$G^{(c)}_{f_1f_0\pi^0 }(s)$ in Eq.~(\ref{Eq3B}) as
\begin{eqnarray} \label{Eq7B} G^{(c)}_{f_1f_0\pi^0
}(s)=m^2_{f_1}\,\bar{g}_{f_1f_0\pi^0}
\,C^{(c)}_{11}/2,\,\end{eqnarray} where the amplitude
\begin{eqnarray} \label{Eq8B} C^{(c)}_{11}=\frac{1}{4m^2_{f_1}p^2(s)}
\Biggl\{C_0(s;m_{K^*},m_{K^+},m_{K^-}) \nonumber\\
\times\Bigl[m^4_\pi-m^2_\pi(m^2_{f_1}+s+m^2_{K^*}-m^2_{K^+})
\nonumber\\ +(m^2_{K^*}-m^2_{K^+})(m^2_{f_1}-s)\Bigl] \nonumber\\
-2m^2_\pi\Bigl[B_0 (m^2_\pi;m_{K^*},m_{K^+})
-B_0(s;m_{K^+},m_{K^-})\Bigl] \nonumber\\ +\Bigl[B_0
(m^2_{f_1};m_{K^*},m_{K^+})-B_0(s;m_{K^+},m_{K^-})\Bigl]\nonumber\\
\times(m^2_{f_1}+m^2_\pi-s)\Biggl\}\,.\ \end{eqnarray} Here,
$C_0(s;m_{K^*},m_{K^+},m_{K^-})$ is the amplitude of the triangle
loop (see Fig.~\ref{Fig3}) in the case of the scalar particles, in
which as the arguments shown are the square of the virtual invariant
mass of the produced  $f_0(980)$ resonance  ($s$) and the masses of
the particles inside the loop,  $B_0 (p^2_i;m_a,m_b)-B_0
(p^2_j;m_c,m_d)$ are the differences of the amplitudes of the
two-point diagrams also for the case of the scalar particles; these
differences are related with the functions $\widetilde{B}_0
(p^2_i;m_a,m_b)$, given in Appendix \ref{appA}, by the relations
$B_0 (p^2_i;m_a,m_b)-B_0 (p^2_j;m_c,m_d)=\widetilde{B}_0
(p^2_i;m_a,m_b)-\widetilde{B}_0
(p^2_j;m_c,m_d)-[\ln(m_am_b/m_cm_d)]/(16\pi^2)$;
$4m^2_{f_1}\,p^2(s)=m^4_{f_1}-2m^2_{f_1}(s+m^2_\pi)+(s-m^2_\pi)^2$.
At points where $1/p^2(s)$ goes to infinity, the function
$C^{(c)}_{11}$ is finite. The numerical evaluation of the amplitude
$C_0(s;m_{K^* },m_{K^+},m_{K^-})$ was fulfilled along the lines
suggested in Ref.~\cite{tHV79}. Notice that the amplitude for the
diagram in Fig.~\ref{Fig3} does not contain the logarithmic
(triangle) singularity in the physical region of the decay
$f_1(1285)\to f_0(980)\pi^0\to\pi^+\pi^-\pi^0$, contrary to the case
of the decay  $\eta(1405)\to f_0(980)\pi^0\to\pi^+ \pi^-\pi^0$,
where it is important to take into account the finite width of the
$K^*(892)$ meson when calculating \cite{AKS15}.

Analogously, for the neutral intermediate states we have
$G^{(n)}_{f_1 f_0\pi^0}(s)=-m^2_{f_1}\bar{g}_{f_1f_0
\pi^0}C^{(n)}_{11}/2$, where $C^{(n)}_{11}$ is obtained from
$C^{(c)}_{11}$ upon the substitution  $m_{K^+(K^-)}$ by
$m_{K^0(\bar K^0)}$.



\begin{thebibliography}{99}
\bibitem{FN1}   When going from Eqs.~(\ref{EqIII-1}) and (\ref{EqIII-2})
to (\ref{EqIII-3}) and (\ref{EqIII-3a}) we do not take into account
the dependence of the kinematic factors entering in the
$J/\psi\to\phi f_0(980)$ and $\chi_{c1}\to a^0_0(980)\pi^0$ vertices
on invariant virtual masses of the scalar mesons in view of its
insignificance in the decays of heavy $J/\psi(3097)$ and
$\chi_{c1}(3511)$ resonances. In this case the corresponding
kinematic factors in the mentioned ratios simply  cancel. Also, we
have neglected the interference of the contributions in the decay
$\chi_{c1}\to a^0_0(980)\pi^0\to \eta\pi^0\pi^0$, related with the
interchange of the indistinguishable $\pi^0 $ mesons. This
interference is small, because the relative momenta of the
$a^0_0(980)$ and $\pi^0$ mesons in the decay  $\chi_{c1}(3510)\to
a^0_0(980)\pi^0$ are large. By the same reason explicitly taking
into account the kinematic factors [$S$ or $D$ wave ones in $J/
\psi\to\phi f_0(980)$  and $P$ wave one in $\chi_{c1}\to a^0_0(980)
\pi^0$] does not imply any sizable deviations from the magnitudes of
coupling constants in Eqs.~ (\ref{EqIII-13a})--(\ref{EqIII-13d})
found by us with the use of Eqs.~(\ref{EqIII-3}) and
(\ref{EqIII-3a}).
\bibitem{FN2} Since the $f_1(1285)$ resonance with the mass
$m_{f_1}=(1281.9\pm0.5)$ MeV \cite{PDG14} is located only by  $160$
MeV above the nominal  production threshold of the $a_0(980)\pi$,
then in the calculations one should of course to take into account
the $P$ wave character of the decay   $f_1(1285)\to a_0(980)\pi$ and
the interference of two amplitudes in the channel   $f_1(1285)\to
a_0(980)\pi\to\eta\pi\pi$ \cite{FN1}. Notice that this interference
is destructive.
\end{thebibliography}
\end{document}